\documentclass[12pt]{article}
\usepackage{psfrag,amsmath,epsfig,feynarts,a4,cite,latexsym}
\usepackage[utf8x]{inputenc}
\usepackage[T1]{fontenc}
\def\pslash#1{{\setbox0=\hbox{$#1$}
  \rlap{\ifdim\wd0>.7em\kern.22\wd0\else\kern.1\wd0\fi /}#1}}
\def\psl{\pslash p}
\def\ksl{\pslash k}
\def\qsl{\pslash q}
\def\Dsl{\pslash D}

\def\sq{{\tilde{q}}}

\def\glui{{\tilde{g}}}
\def\epsbar{{\bar{\epsilon}}}

\def\La{{\cal L}}
\def\Layu{\La_{y_u}}

\newcommand{\gluibar}{\overline{\tilde{g}}}
\newcommand{\qbar}{\bar{q}}

\newcommand{\ygluibar}{{\overline{\tilde{y}_{\glui}}}{}}

\allowdisplaybreaks[1]
\usepackage{booktabs}
\usepackage{multirow}

\begin{document}
%\begin{flushright}
%\end{flushright}
%\vspace{3em}
\begin{center}
{\Large\bf Three-loop MSSM Higgs-Boson Mass Predictions 
}\\[2ex]
{\Large\bf and Regularization by Dimensional Reduction
}
\\
\vspace{3em}
%{\large {\renewcommand{\thefootnote}{\fnsymbol{footnote}}
%\footnote{\parbox[t]{10cm}{}}}}
%  \\[2ex]
%  \parbox{10cm}{%\small\center\em
{\sc Dominik St\"ockinger, Josua Unger$^1$
}\\[2em]
{\it ${}^1$\ Institut f\"ur Kern- und Teilchenphysik\\
Technische
  Universit\"at Dresden, Germany
}
\setcounter{footnote}{0}
\end{center}
\vspace{2ex}
\begin{abstract}

The evaluation of three-loop contributions to the MSSM
Higgs-boson mass is considered at the orders enhanced by the strong
gauge coupling and top or bottom Yukawa couplings, i.e.\ at the orders
${\cal 
  O}(\alpha_{t,b}\alpha_s^2,\alpha_{t,b}^2\alpha_s,\alpha_{t,b}^3)$. We
prove that  
regularization by dimensional reduction preserves supersymmetry at the
required level. Thus generating counterterms by multiplicative
renormalization is correct. Technically, we extend a previous two-loop
analysis to the three-loop level. The extension covers not only the
genuine three-loop Higgs potential counterterms but also a large
sector of two-loop counterterms, required for subrenormalization.

\end{abstract}

\section{Introduction}

After the discovery of the Higgs boson at the LHC
\cite{LHCHiggs}, the Higgs boson mass $M_h$ has
become a precision observable. Supersymmetric (SUSY) extensions of the
Standard Model are special because they allow to predict the Higgs
boson mass thanks to SUSY relations between the Higgs potential and
known gauge couplings. Specifically in the MSSM, $M_h$ is smaller than
the Z boson mass at tree level, and calculable higher-order
corrections can push it up to the measured value, resulting in
important constraints on SUSY parameters.

Precision evaluations of these MSSM higher-order corrections have a
long history. One-loop and leading 2-loop corrections have been
evaluated long ago; we refer to the recent review
\cite{Draper:2016pys} for details and references. The leading
corrections are governed by Yukawa couplings $\alpha_{t,b}$ and the
strong gauge coupling $\alpha_s$. They can be evaluated in the
so-called gaugeless limit, where the electroweak gauge couplings and
$M_{W,Z}$ go to zero with fixed ratio $M_W/M_Z$ and fixed Higgs vacuum
expectation values. At the 3-loop level, the full corrections of order ${\cal
  O}(\alpha_t\alpha_s^2)$
%notation from Draper,Rzehak
have been evaluated in the gaugeless limit
in Refs.\  \cite{HarlanderKantHiggs} and further developed
in Refs.\ \cite{Kunz:2014gya,Harlander:2017kuc}. Leading and
subleading 3-loop large logarithms have been evaluated in
Ref.\ \cite{Martin:2007pg}.  Further recent
progress includes 2-loop computations beyond the gaugeless limit and
the resulting zero-momentum approximation
  \cite{Borowka:2014wlaetal,Degrassi:2014pfa,Borowka:2018anu},
  resummation of large logarithms using EFT and renormalization 
group methods
\cite{eftresults,Bagnaschi:2017xid}
and the development of hybrid approaches
\cite{hybridFH,Athron:2016fuq}. The
remaining theory uncertainties of these current calculations depends
on the details of the spectrum and on the calculational scheme, see
e.g.\ the recent discussions in
Refs.\ \cite{Athron:2016fuq,Bagnaschi:2017xid}. However, the
theoretical uncertainty is significantly larger than the experimental
one --- hence it is motivated to further improve the accuracy of
theoretical calculations.

Here we consider the technically necessary step of regularization of
higher-order loop contributions. The commonly used scheme is
regularization by dimensional reduction (DRED) \cite{Siegel79,CJN80} (see
also the recent review \cite{Gnendiger:2017pys} of regularization
schemes). This scheme is consistent with SUSY for the 2-loop
calculations in the gaugeless limit \cite{HollikDS05}. We ask whether
the same is true at the 3-loop level. If this turned out not to be the
case, additional 
SUSY-restoring counterterms would be required, which would affect the
finite, non-logarithmic parts of a 3-loop calculation, i.e.\ the parts
which cannot be obtained by renormalization-group methods.

In the remainder of this Introduction, we provide more details on
potential non-symmetric counterterms and DRED.

In a 3-loop calculation of the Higgs boson masses in the gaugeless
limit of the MSSM several
types of counterterms are required: 
\begin{itemize}
\item genuine 3-loop counterterms in the
Higgs sector;
\item 1-loop and 2-loop counterterms in the Higgs, Yukawa,
and SUSY-QCD sectors. 
\end{itemize}
Usually, the structure of the counterterm Lagrangian and Feynman rules
is generated by a renormalization transformation respecting the
symmetries of the theory. In this way the generated
counterterms respect in particular SUSY. The present paper
investigates whether this is correct and 
sufficient for calculations of MSSM Higgs boson masses at the 3-loop level
in the gaugeless limit.

Cases in which a renormalization transformation respecting SUSY is not
enough and additional, non-SUSY counterterms are needed are familiar
in dimensional regularization (DREG). DREG breaks SUSY already at the
1-loop level, and non-SUSY counterterms have to be added e.g.~to the
gluino--squark--quark interaction. Such non-SUSY counterterms required
in DREG have been evaluated and documented in
Refs.~\cite{MartinVaughn,Mihaila:2009bn,Stockinger:2011gp}. 

For DRED, on the other hand, many studies have confirmed the
compatibility with SUSY and the absence of non-SUSY counterterms, for
overviews of earlier results see \cite{JJ,DS05,HollikDS05}. In
particular, several 
2-loop cases of relevance for us have been studied: self energies and
SUSY transformations of chiral multiplets \cite{DS05} in general SUSY
gauge theories, quartic Higgs coupling in the gaugeless limit of the
MSSM \cite{HollikDS05}, three-point interactions in the gauge/gaugino
sector of SUSY Yang-Mills theories \cite{Harlander:2006xq} and in the
gauge/gaugino/quark/squark sector of SUSY QCD
\cite{Harlander:2009mn}. The existing studies do not 
cover the case of potential, finite non-SUSY counterterms to Higgs
masses at the 3-loop level.

Also beyond the question of SUSY preservation, the status of DRED is
similarly mature as the one of dimensional regularization: Both
schemes can be formulated in a mathematically consistent way and the
quantum action principle holds
\cite{BM,DS05}, and the two schemes are equivalent \cite{JJR},
i.e.\ one can always find local transition counterterms which
translate between the two schemes. The infrared divergence structure
of DRED regularized QCD amplitudes has been understood
\cite{IRstructure}.
All these statements require to
distinguish the formally $D$-dimensional parts of vector fields and
the remaining $(4-D)$-dimensional parts (the so-called
$\epsilon$-scalars), and in particular to treat the renormalization
and factorization of
$\epsilon$-scalars as independent. The need for the
independent renormalization has also been stressed in concrete
multi-loop calculations
\cite{multiloopepsilonscalars}.

The recent progress in understanding regularizations and the need for
more precise Higgs boson mass computations motivates our study. The
outline of the remainder of the paper is as follows.
In section \ref{sec:setup} we describe the setup and notation, and we
discuss the different structure of SUSY and 
non-SUSY counterterms. We also explain the methods used to
determine potential non-SUSY counterterms. Section
\ref{sec:lowerorder} is devoted to the analysis of 2-loop
counterterms, which are required for subrenormalization and shows that
DRED is consistent with SUSY on this level. Section
\ref{sec:threeloop} finally discusses the proof that no SUSY-restoring
counterterm is required for the MSSM Higgs boson mass calculations at
the 3-loop level in the gaugeless limit.

\section{Setup and SUSY versus non-SUSY counterterms}
\label{sec:setup}

\subsection{Setup}

In this subsection we describe our setup, provide
relevant notation and discuss the structure of the usual symmetric
counterterms. We work in the MSSM with neglected electroweak gauge 
couplings, $g_{1,2}\to0$, the so-called gaugeless limit. The
relevant fundamental fields are the Higgs and Higgsino doublets
${H}_k^i$, 
$\tilde{H}_k^i$ (where $k=1,2$ and $i$ is the SU(2) index and where
the vacuum expectation values have already been split off), the
left-handed quark and squark doublets $q^i_L$ and $\tilde{q}_L^i$,
and the right-handed quark and squark singlets $u_R$, $d_R$,
$\tilde{u}_R$, $\tilde{d}_R$, as well as the gluon and gluino fields
$A^\mu$, $\tilde{g}$. All spinors are taken as 4-spinors; for the
quarks we abbreviate $q_L=P_L q$, $u_R=P_Ru$ etc; the gluino
spinor is a Majorana spinor; the Higgsino spinors are taken as purely
left-handed, $\tilde{H}_k^i\equiv\tilde{H}_k^i{}_L=P_L\tilde{H}_k^i$. Gluon and gluino fields
without SU(3) index are taken as contracted with Gell-Mann matrices,
$\tilde{g}=\tilde{g}^a\frac{\lambda^a}{2}$, etc, and (s)quark colour
indices are always suppressed.

The Lagrangian containing the dimension-4-terms relevant for our analysis
can be decomposed as
\begin{align}
  \La &=
  \La_{\text{kin}}+\Layu+\La_{\text{gluino}}-V_{\text{quartic}}\, .
\end{align}
Here $\La_{\text{kin}}$ contains the usual kinetic terms for all
fields with the SU(3) gauge covariant derivative,
\begin{align}
  \La_{\text{kin}} &= \sum_{q=u,d}\bar{q}i\Dsl
  q+\sum_{k=1,2}|\partial^\mu H_k|^2 + {\text{kinetic terms for
  }}\tilde{q},\tilde{H},A^\mu,\tilde{g},
  \\
  D^\mu&=\partial^\mu+ig_sA^\mu .
\end{align}
The remaining terms are the up-type Yukawa interactions, gluino
interactions and the quartic Higgs potential:
\begin{align}
  \Layu &= y_u \epsilon_{ij}\left(
  H_2^i \bar{u} q^j_L
  + \tilde{u}_R^\dagger \overline{\tilde{H}^C_2}^i q^j_L
  + \bar{u}  \tilde{H}^i_2{}_L\tilde{q}_L^j
  \right)+ h.c.,\\
  \La_{\text{gluino}} &=  -\sqrt2g_s\sum_{q=u,d}\left(\sq_L^\dagger \gluibar  q_L + \qbar \glui_R \sq_L
              -\sq_R^\dagger \gluibar  q_R - \qbar \glui_L \sq_R
              \right),\\
  V_{\rm quartic} &=\frac{g_1^2+g_2^2}{8}
\left(\left|{H}_1\right|^2-\left|{H}_2\right|^2\right)^2
 + \frac{g_2^2}{2}\left|{H}_1^\dagger {H}_2\right|^2 \,.
\end{align}
In the last term the electroweak gauge couplings $g_1,g_2$ have been
retained for reference. We have only considered up-type
Yukawa couplings in $\Layu$ and suppressed corresponding down-type
Yukawa coupling terms because they can be treated in the same way. We
have also suppressed quartic terms involving squarks and Lagrangian
terms of dimension $\le3$ and soft SUSY 
breaking terms since they are irrelevant for our goal of determining
potential SUSY breaking in the dimension-4 sector of the Higgs
potential.

The basic symmetries of the MSSM are Lorentz and gauge invariance and (softly
broken) SUSY. The MSSM further respects a softly broken Peccei-Quinn
symmetry, which forbids a dimension-4 mixing between the two Higgs
doublets. The symmetries can be encoded in Slavnov-Taylor and Ward
identities, allowing a full study of the renormalization of the MSSM
\cite{HKRRSS} and providing the most general form of symmetric
counterterms, i.e.\ the correct counterterm structure for the
desirable case where the regularization
preserves the symmetries of the theory. Specializing the results of
Ref.\ \cite{HKRRSS} for the full MSSM to the case at hand we obtain
the following multiplicative renormalization transformation of the
parameters and fields:
\begin{subequations}
  \label{SymRenTrans}
  \begin{align}
  g_s &\to Z_{g_s}g_s,\\
  y_u &\to Z_{y_u}y_u,\\
  \varphi &\to \sqrt{Z_\varphi}\varphi\text{ for all fields
  }\varphi=u_{L,R},d_{L,R},\tilde{u}_{L,R},\tilde{d}_{L,R},
          {H}_k,\tilde{H}_k,A^\mu,\tilde{g},
          \\
  Y_\varphi &\to \sqrt{Z_\varphi}^{-1} Y_\varphi \,.
\end{align}
\end{subequations}
In the last of these equations we have also specified the
renormalization transformation for sources of BRS transformations,
which will appear later in the Slavnov-Taylor identity.

Applying this symmetric renormalization transformation to the
classical action $\Gamma_{\text{cl}}$ generates  a
bare action (i.e.\ the sum of classical and counterterm action)
\begin{align}
  \Gamma_{\text{cl}}\stackrel{\text{(\ref{SymRenTrans})}}{\longrightarrow}
  \Gamma_{\text{cl+ct-sym}}
  \,,
\end{align}
which contains the usual symmetric counterterms.

For our purposes, the decisive features of the symmetry-preserving nature of the
resulting counterterms $\Gamma_{\text{ct-sym}}$ are that the renormalization of all three terms
in $\Layu$ and the interaction terms involving gluons
or gluinos are correlated, i.e.\ that the SUSY-relations between
interactions of particles and superpartners are respected. Furthermore,
the resulting counterterm for 
$V_{\text{quartic}}$ vanishes in the gaugeless limit. Later we will
compare this form of the counterterms to a more general one, which
does not respect SUSY.

\subsection{Method to determine potential non-SUSY counterterms}

The basic idea to compute or check the absence of non-SUSY
counterterms is common to all mentioned studies in the literature. One
needs to evaluate 
quantities which depend on the potential non-SUSY counterterms and
which satisfy known relations. Relations employed in the literature
include
\begin{itemize}
\item equality of on-shell amplitudes in the exact SUSY limit;
\item equality of $\beta$ functions at higher orders;
\item validity of SUSY Slavnov-Taylor identities.
\end{itemize}
In particular Refs.~\cite{Harlander:2006xq,Harlander:2009mn} used the
second approach and computed 
3-loop $\beta$ functions for various gluon, gluino, and
$\epsilon$-scalar couplings. These computations depend on finite
2-loop counterterms, and it was shown that the SUSY relations between
the $\beta$ functions are satisfied if finite, non-SUSY counterterms
are {\em absent} at the 2-loop level.

Now we give a brief overview of our approach, which is the one of
Refs.\ \cite{STIChecks2,STIChecks3,HollikDS05} and based on
Slavnov-Taylor identities. The first 
ingredient is the Slavnov-Taylor identity (STI) expressing SUSY and gauge
invariance and the respective (anti-)commutation relations on the
level of Green functions. It can be written as
\begin{align}
  S(\Gamma)=0,
\end{align}
where $\Gamma$ is the generating functional of
one-particle irreducible (1PI) Green functions, and where 
\begin{align}
S(\Gamma) &=
\int d^4x \left(
\frac{\delta\Gamma}{\delta Y_{\varphi_i}(x)}\frac{\delta\Gamma}{\delta
  \varphi_i(x)}
+s\varphi_i'(x) \frac{\delta\Gamma}{\delta \varphi_i'(x)}
\right)
\,.
\end{align}
Here $\varphi_i$ runs over the MSSM quantum fields with nonlinear BRS
transformations $s\varphi_i$, and $\varphi_i'$ runs over the MSSM
quantum fields with linear BRS transformations $s\varphi_i'$.
The  BRS transformations contain in particular the SUSY
transformations, where the SUSY ghost $\epsilon$ acts as the
transformation parameter.
The $Y_{\varphi_i}$ are sources of the nonlinear BRS transformations, i.e.\ the
classical Lagrangian contains a part ${\cal L}_{\text{ext}}=Y_{\varphi_i}
s\varphi_i$. The explicit form of the MSSM STI identity can
be found in Ref.~\cite{HKRRSS}, and adaptations to 4-spinor notation
are given in Refs.\ \cite{STIChecks2,STIChecks3}.

The STI has to hold {\em after} renormalization, i.e.\ after taking
into account counterterms of all orders. Evaluating the STI at the
$n$-loop level thus constrains the genuine $n$-loop counterterms in
terms of regularized (and subrenormalized) $\le n$-loop diagrams.
In the case that the STI is
already valid on the regularized/subrenormalized level, the non-SUSY
counterterms are zero and the usual renormalization transformation
Eq.\ (\ref{SymRenTrans})
(the full form of which has been given in Ref.~\cite{HKRRSS}) is
sufficient.

Specific STIs for concrete sets of Green functions can be obtained by
taking functional derivatives of the equation $S(\Gamma)=0$ with
respect to sets of fields. Such specific STIs can then be used to
determine potential non-SUSY counterterms in specific sectors of the
MSSM. 

The other ingredient of our approach is the regularized quantum action
principle in DRED \cite{DS05}. It provides a direct way to check the
validity of STIs on the regularized level and thus to check the
absence of non-SUSY counterterms. Applied to the STI it can be written as 
\begin{align}
S(\Gamma^{\text{DRED}}) &=
i[\Delta]\cdot \Gamma^{\rm DRED},
&
\Delta &\equiv S(\Gamma_{\text{cl+ct}})
\label{STIQAP}
\end{align}
where all quantities are regularized in DRED and taken in $D$
dimensions. The notation for such regularized quantities is as
follows. $\Gamma_{\text{cl+ct}}$ is the 
bare action which is used to define the Feynman rules of
the regularized (and partially or fully renormalized)
theory. $\Gamma^{\rm {DRED}}$ is the resulting generating functional
for 1PI Green functions.
$[\Delta]\cdot \Gamma^{\rm DRED}$ denotes the insertion of 
the operator $\Delta$ into the 1PI Green functions; the operator is
defined via the STI applied to the bare action (the sum of classical
and counterterm action).
The general result for the lowest-order term
$\Delta_{\text{cl}}\equiv S(\Gamma_{\text{cl}})$
has been given in Ref.~\cite{DS05}. This operator
is non-vanishing, corresponding to the fact that in $D$ dimensions
even in DRED, Fierz identities are invalid and therefore the classical
action cannot be shown to be supersymmetric. 

In the following two sections we will use the STI and the quantum
action principle to determine first the structure of 2-loop
counterterms in all sectors and then the 3-loop counterterms for the
Higgs mass calculation.

\section{Lower-order results}
\label{sec:lowerorder}

In this section we determine the potential non-SUSY counterterms at
2-loop order in all relevant sectors of the MSSM in the gaugeless
limit. These counterterms are important for two reasons. On the one
hand they would appear in explicit computations of $M_h$ at the
3-loop level such as Refs.~\cite{HarlanderKantHiggs} or
future extensions thereof; on the 
other hand they would appear in the computation of $\Delta$ in the
quantum action principle Eq.~(\ref{STIQAP}) needed to evaluate the 3-loop
STI below in section \ref{sec:threeloop}.

The required structure of 2-loop counterterms is already essentially
known. Ref.~\cite{DS05} has considered self energies and SUSY
transformations of chiral multiplets;
Refs.~\cite{Harlander:2006xq,Harlander:2009mn} have considered 
3-point interactions between (s)quarks and gluons, gluinos and
$\epsilon$-scalars using the method of higher-order $\beta$
functions. The result of all these references is that non-SUSY
counterterms are not needed and the renormalization transformation
respecting SUSY gives the correct counterterm structure.

Here we slightly extend these results, in order to
illustrate and confirm the method of the STI and quantum action
principle. We focus on counterterms involving
interactions between Higgs/Higgsino and (s)quarks, and terms involving
the BRS sources $Y_{\varphi_i}$ and SUSY ghosts $\epsilon$. We remark that
2-loop counterterms to 4-point self interactions of squarks or
$\epsilon$-scalars are not important for our purposes and not covered
by either the analysis of
Ref.~\cite{Harlander:2006xq,Harlander:2009mn} or the following
analysis. 

We begin in subsection \ref{sec:exemplarySTI} with a detailed
discussion of an exemplary STI; later in subsection \ref{sec:otherSTI}
we will be briefer.

\subsection{Exemplary case}
\label{sec:exemplarySTI}

Here we explain in detail the derivation of the 2-loop counterterm
structure in the sector of the Yukawa interactions between
Higgs/Higgsino and up-(s)quarks. We will show that the symmetric
counterterms generated by Eq.\ (\ref{SymRenTrans}) are correct in the context of
DRED. To set the stage we first provide the form of the most general
bare Lagrangian of this sector, which is {\em not} restricted by
SUSY, only by manifest symmetries of DRED (such as Lorentz and gauge
invariance and (softly broken) Peccei-Quinn symmetry). It reads
\begin{align}
\label{Layubare}
  \Layu^{\text{cl+ct, general}} &=
   y_u \epsilon_{ij}\left( Z_{H_2q{u}}
  H_2^i \bar{u} q^j_L
  + Z_{\tilde{H}_2q\tilde{u}}\tilde{u}_R^\dagger \overline{\tilde{H}^C_2}^i  q^j_L
  + Z_{\tilde{H}_2\tilde{q}u} \bar{u} \tilde{H}^i_2{}_L\tilde{q}_L^j
  \right)+ h.c.\, .
\end{align}
It contains the same terms as the tree-level Lagrangian $\Layu$, but
each term appears multiplied with an independent coefficient $Z_{i}$.

Our claim is that symmetric counterterms are correct in this
sector. Equivalently we need to show that the following special choice
for the $Z_i$, corresponding to the symmetric renormalization 
transformation (\ref{SymRenTrans}), is correct:
\begin{subequations}
  \label{claimyuksector}
  \begin{align}
\label{claimyukDef}  Z_{H_2q{u}} &= Z_{y_u}\sqrt{Z_{H_2} Z_{q_L} Z_{u_R}},\\
  Z_{\tilde{H}_2q\tilde{u}} &= Z_{y_u}\sqrt{Z_{\tilde{H}_2} Z_{q_L} Z_{\tilde{u}_R}},\\
  Z_{\tilde{H}_2\tilde{q}{u}} &= Z_{y_u}\sqrt{Z_{\tilde{H}_2} Z_{\tilde{q}_L} Z_{u_R}}.
\end{align}
\end{subequations}
The first of these three equations can be assumed to hold by
definition --- it defines the quantity $Z_{y_u}$; the other two
equations are then nontrivial. The physical meaning
is that there is only one fundamental
Yukawa coupling parameter, which governs all three interactions.

In order to prove the second equation of (\ref{claimyuksector}), we
consider the functional derivative of the STI
\begin{align}
  \label{STIexample}
  0&=\frac{\delta^4 S(\Gamma)}{\delta q_L^i \delta \tilde{u}_R^\dagger
    \delta H_2^j\delta \bar\epsilon}\,.
\end{align}
Here $\epsilon$, the SUSY ghost, is a bosonic Majorana spinor. The
equation thus corresponds to a SUSY relation connecting $q_L$, $H_2$,
$\tilde{u}_R$. 
More details to the use, derivation and evaluation of such identities can be
found in Refs.\ \cite{STIChecks2,STIChecks3}. Like in Ref.\ \cite{STIChecks2} we will
use the identity only at leading order in the external momenta, so
that contributions from soft supersymmetry breaking or electroweak
symmetry breaking can be neglected.
Evaluating Eq.\ (\ref{STIexample}) yields the following STI between 1PI Green functions,
\begin{align}
  \label{STIqsqHfull}
  0=&
  - \Gamma_{\tilde{u}_R^\dagger\,\bar\epsilon \, y_{u_R}}
  \Gamma_{q_L^i \,H_2^j \,\bar{u}_R}
  - \Gamma_{H_2^j \,\bar\epsilon \,y_{\tilde{H}_l^m}{}^C}
  \Gamma_{q_L^i \,\tilde{u}_R^\dagger \,\overline{\tilde{H}_l^m}^C}
  - \Gamma_{H_2^j \,\tilde{u}_R^\dagger \,\bar\epsilon \,y_{q_L^l}}
  \Gamma_{q_L^i \, \bar{q}_L^l }
  \nonumber\\
  &
  + \Gamma_{q_L^i \,\bar\epsilon \,Y_{\tilde{q}_L^l}}
  \Gamma_{    \tilde{u}_R^\dagger \,H_2^j \, \tilde{q}_L^l}
  + \Gamma_{q_L^i \,\tilde{u}_R^\dagger \,\bar\epsilon \,Y_{\phi_l}}
  \Gamma_{H_2^j \,\phi_l}
  + \Gamma_{q_L^i \,H_2^j \,\bar\epsilon \,Y_{\tilde{u}_R}}\Gamma_{
    \tilde{u}_R^\dagger \,\tilde{u}_R}
  \,.
\end{align}
Generically, the symbols $Y_{\varphi_i}$ 
denote the sources of the BRS variations of the respective fields
$\varphi_i$, where lower-case notation $y_{\varphi_i}$ is used for 4-spinor fields;
$\phi_l$ runs over all Higgs field components.

Each of the six terms is a product of one elementary 1PI Green
function and one 1PI Green function with an insertion of a BRS
transformation. The three terms in the second line can be ignored in
the following: the last two terms in the second line involve
Green functions which do not 
receive tree-level or counterterm contributions. The first term in the
second line is subleading in the external momenta due to its Lorentz
structure. Furthermore, we will now specialize to the case in which
the quark field $q_L^i$ carries no momentum, $p_{q_L}=0$. Then also the
last term in the first line can be neglected, and the identity
simplifies to
\begin{align}
\label{STIqsqH}
  0&=
  - \Gamma_{\tilde{u}_R^\dagger \, \bar\epsilon \, y_{u_R}}
  \Gamma_{q_L^i \, H_2^j \,\bar{u}_R}
  - \Gamma_{H_2^j \, \bar\epsilon \, y_{\tilde{H}_l^m}{}^C}
  \Gamma_{q_L^i \, \tilde{u}_R^\dagger \, \overline{\tilde{H}_l^m}^C} +\ldots
\end{align}
where the dots denote the neglected terms.

This is an identity between the first and second kind of the Yukawa
interactions in $\Layu$. The Green functions appearing as prefactors
correspond to loop-corrected SUSY transformations of $u_R$ into
$\tilde{u}_R$ and of $\tilde{H}_2$ into $H_l$, respectively.

\begin{figure}[tb]
\begin{center}
\unitlength=0.9cm%
\begin{feynartspicture}(16,4)(4,1)
\FADiagram{(a)}
\FAVert(10,16){1}
\FALabel(3,20)[r]{$\epsbar$\ }
\FAProp(17,20)(10,16)(0.,){/Straight}{0}
\FALabel(17,20)[l]{\ ${q_L}$}
\FAProp(3,20)(10,16)(0.,){/Straight}{0}
\FALabel(4,14)[r]{$q$}
\FALabel(16,14)[l]{$\tilde{H},\glui$}
\FALabel(16,7)[l]{$q$}
\FALabel(4,7)[r]{$\tilde{H}$}
\FALabel(10,4.5)[b]{$q$}
\FALabel(10,10.5)[b]{$\tilde{q}$}
\FAProp(4,10)(16,10)(0.,){/ScalarDash}{0}
\FAProp(10,16)(10,4)(1.,){/Straight}{0}
\FAProp(10,16)(10,4)(-1.,){/Straight}{0}
\FAVert(5.5,6){0}
\FAVert(14.5,6){0}
\FAProp(5.5,6.)(2,1)(0.,){/ScalarDash}{0}
\FAProp(18,1)(14.5,6.)(0.,){/ScalarDash}{0}
\FALabel(0,0)[l]{$\tilde{u}_R$}
\FALabel(20,0)[r]{$H$}
\FAVert(4,10){0}
\FAVert(16,10){0}
\FADiagram{(b)}
\FAVert(10,16){1}
\FALabel(3,20)[r]{$\epsbar$\ }
\FAProp(17,20)(10,16)(0.,){/Straight}{0}
\FALabel(17,20)[l]{\ ${q_L}$}
\FAProp(3,20)(10,16)(0.,){/Straight}{0}
\FALabel(4,14)[r]{$\tilde{H}$}
\FALabel(16,14)[l]{$q$}
\FALabel(16,7)[l]{$q$}
%\FALabel(4,7)[r]{$q$}
\FALabel(8,4.5)[b]{$q$}
\FALabel(11,9)[b]{$g$}
\FAProp(16,10)(10,4)(0.5,){/Cycles}{0}
\FAProp(10,16)(10,4)(1.,){/Straight}{0}
\FAProp(10,16)(10,4)(-1.,){/Straight}{0}
\FAVert(5.5,6){0}
\FAVert(14.5,6){0}
\FAProp(5.5,6.)(2,1)(0.,){/ScalarDash}{0}
\FAProp(18,1)(14.5,6.)(0.,){/ScalarDash}{0}
\FALabel(0,0)[l]{$\tilde{u}_R$}
\FALabel(20,0)[r]{$H$}
\FAVert(16,10){0}
\FAVert(10,4){0}
%% \FADiagram{(c)}
%% \FAVert(4,10){0}
%% \FAVert(16,10){0}
%% \FAVert(10,16){1}
%% \FALabel(3,20)[r]{$\epsbar$\ }
%% \FAProp(17,20)(10,16)(0.,){/Straight}{0}
%% \FALabel(17,20)[l]{\ ${\tilde{H}}$}
%% \FAProp(3,20)(10,16)(0.,){/Straight}{0}
%% \FALabel(4,14)[r]{$q$}
%% \FALabel(16,14)[l]{$q$}
%% \FALabel(16,6)[l]{$q$}
%% \FALabel(10,9.5)[t]{$g$}
%% \FAProp(4,10)(16,10)(0.,){/Cycles}{0}
%% \FAProp(10,16)(10,4)(1.,){/Straight}{0}
%% \FAProp(10,16)(10,4)(-1.,){/Straight}{0}
\FADiagram{(c)}
\FAVert(10,16){1}
\FALabel(3,20)[r]{$\epsbar$\ }
\FAProp(17,20)(10,16)(0.,){/Straight}{0}
\FALabel(17,20)[l]{\ ${q_L}$}
\FAProp(3,20)(10,16)(0.,){/Straight}{0}
\FALabel(4,14)[r]{$\tilde{H}$}
\FALabel(18,14)[l]{\,$g$}
\FALabel(16,7)[l]{$q$}
%\FALabel(4,7)[r]{$q$}
\FALabel(10,4.5)[b]{$q$}
\FAProp(16,10)(14.66,18.66)(0.5,){/Cycles}{0}
\FAProp(10,16)(10,4)(1.,){/Straight}{0}
\FAProp(10,16)(10,4)(-1.,){/Straight}{0}
\FAVert(5.5,6){0}
\FAVert(14.5,6){0}
\FAProp(5.5,6.)(2,1)(0.,){/ScalarDash}{0}
\FAProp(18,1)(14.5,6.)(0.,){/ScalarDash}{0}
\FALabel(0,0)[l]{$\tilde{u}_R$}
\FALabel(20,0)[r]{$H$}
\FAVert(16,10){0}
\FAVert(14.66,18.66){0}
\FADiagram{(d)}
\FAVert(14.66,18.66){0}
\FAVert(16,10){0}
\FAVert(10,16){1}
\FALabel(3,20)[r]{$\epsbar$\ }
\FAProp(17,20)(10,16)(0.,){/Straight}{0}
\FALabel(17,20)[l]{\,$q_L$}
\FAProp(3,20)(10,16)(0.,){/Straight}{0}
\FALabel(5,14)[r]{$\glui,q$}
\FALabel(18,14)[l]{\,$\tilde{q}$}
\FALabel(14,13)[r]{$\glui,q$}
\FALabel(12,18)[b]{$\glui$}
\FAProp(10,16)(10,4)(1.,){/Straight}{0}
\FAProp(10,16)(10,4)(-1.,){/Straight}{0}
\FAProp(14.66,18.66)(16,10)(-.5,){/ScalarDash}{0}
\FAVert(5.5,6){0}
\FAVert(14.5,6){0}
\FAProp(5.5,6.)(2,1)(0.,){/ScalarDash}{0}
\FAProp(18,1)(14.5,6.)(0.,){/ScalarDash}{0}
\FALabel(20,0)[r]{$H,\tilde{u}_R$}
\FALabel(0,0)[l]{$\tilde{u}_R,H$}
\FALabel(10,4.5)[b]{$q$}
\end{feynartspicture}
\end{center}
\vspace{-.5cm}
\caption{Diagrams contributing to~(\ref{breakingexample}) at the two-loop
  level in the gauge-less limit. The insertion of the operator
  $\Delta^{\le\text{1L}}\equiv
S(\Gamma_{\text{cl+ct}}^{\le\text{1L}})$ is marked by a cross. Quarks, gluons and Higgs
  bosons are denoted by $q$, $g$, $H$; squarks and
  Higgsinos are denoted by $\tilde{q}$ and $\tilde{H}$. 
 % In the text, the $\gamma$-string attached to the second
 % external fermion line is denoted as $A$, the $\gamma$-string
 % attached to the closed fermion loop as $B$.
}
\label{fig:breakingexample}
\end{figure}

As a first step we have to check whether the STI (\ref{STIexample})
is valid on the regularized 2-loop level with 1-loop
subrenormalization. According to the quantum action principle the
potential violation is given by
\begin{align}
\frac{\delta^4 S(\Gamma^{\rm DRED})}{\delta q_L^i \delta \tilde{u}_R^\dagger
  \delta H_2^j\delta \bar\epsilon}
=&
\left(\left[\Delta^{\le\text{1L}}\right]\cdot\Gamma^{\rm DRED}\right)_{ q_L^i \,
  \tilde{u}_R^\dagger \, H_2^j \, \bar\epsilon}
\label{breakingexample}
\end{align}
i.e.~by 1PI diagrams with external fields $q_L^i \,
  \tilde{u}_R^\dagger \, H_2^j \, \bar\epsilon$ and one
insertion of the operator $\Delta^{\le\text{1L}}\equiv
S(\Gamma_{\text{cl+ct}}^{\le\text{1L}})$. Here
$\Gamma_{\text{cl+ct}}^{\le\text{1L}}$ is the 1-loop 
bare action. At this level DRED is clearly SUSY-preserving, and
$\Gamma_{\text{cl+ct}}^{\le\text{1L}}$ is obtained from the classical action by the
usual renormalization transformation. Hence also $\Delta^{\le\text{1L}}$ at this
order is obtained from $\Delta_{\text{cl}}\equiv S(\Gamma_{\text{cl}})$ given in
Ref.~\cite{DS05} simply by the renormalization transformation, and the
structure of $\Delta^{\le\text{1L}}$ and its Feynman rules are the ones of
Ref.~\cite{DS05}.

Fig.~\ref{fig:breakingexample} shows representative Feynman diagrams
contributing to Eq.~(\ref{breakingexample}). All diagrams have a
common structure with one open fermion line and one closed fermion
loop. As in Refs.~\cite{DS05,HollikDS05} we denote the chain of $\gamma$-matrices
corresponding to the open fermion line as $A$, and the $\gamma$-chain
corresponding to the closed fermion loop as $B$. In the terminology of
those references the first three diagrams of Fig.~\ref{fig:breakingexample} are
of topology (c), and the last diagram is of topology (a) or (b),
in case of gluinos/quarks in the fermion loop. According to the rules
given in that reference, the diagrams vanish if the $\gamma$ chain $B$
can be expressed as a product of up to 4, 2, 3 $\gamma$-matrices (for
topology (a), (b), (c), respectively). These three topologies and their
properties are summarized in Tab.~\ref{tab:topologies}.

\begin{table}
  \centerline{
    \begin{tabular}{c|c|c}
  Topology & Fermions at $A/B/B$ & $\#$ allowed $\gamma$-matrices  \\
   \hline
   Topology (a) & $\glui/\glui/\glui$ & $\le4$ \\
   Topology (b) & $\glui/\psi/\psi$ & $\le2$  \\
   Topology (c) & $\psi/\psi/\psi$ or $\psi/\psi/\glui$ & $\le3$  \\
    \end{tabular}
    }
  \caption{\label{tab:topologies}
  Properties of the three topologies defined in Ref.~\cite{DS05} for
  the diagrams involving an 
  insertion of the operator $\Delta\equiv S(\Gamma_{\text{cl}})$
  attached to an open fermion line  ($A$) and one fermion loop
  ($B$). The second column specifies whether the fermions in line $A$
  or $B$ coupling to the insertion are gauginos (denoted by $\glui$)
  or  fermions of a chiral SUSY multiplet ($\psi$). The
  last column specifies the maximum number of ``allowed''
  $\gamma$-matrices in the $\gamma$-string $B$. If the respective
  bound is satisfied, the diagram vanishes \cite{DS05}.}
\end{table}
After integrating over the fermion-loop momentum, the fermion loop can
depend on up to two momenta (the second loop momentum and the single
non-vanishing external momentum). Hence the fermion loop can be
expressed as a product of up to two (three) gamma matrices in diagrams
of Fig.~\ref{fig:breakingexample} (a,b,d)
(Fig.~\ref{fig:breakingexample} (c)). According to the rule mentioned
above, this is sufficient to know that all diagrams of
Fig.\ \ref{fig:breakingexample} vanish. It is easy to see that the
same is true for all other diagrams which could contribute to the breaking
Eq.\ (\ref{breakingexample}). Hence Eq.\ (\ref{breakingexample})
vanishes, and the STI (\ref{STIqsqH}) is valid at the regularized
2-loop order with 1-loop subrenormalization.

The important consequence of this is that the genuine 2-loop
counterterm contributions have to fulfill the STI
(\ref{STIqsqH}) by themselves. This implies the following identities
of renormalization constants (at the 2-loop level in the gaugeless limit):
\begin{align}
  \label{resultSTI}
  Z_{\tilde{u}\bar\epsilon y_{u_R}} Z_{H_2q{u}} &= Z_{H_2\bar\epsilon
    y_2} Z_{\tilde{H}_2q\tilde{u}}
  \, ,
\end{align}
where we have used the notation of the bare Lagrangian
(\ref{Layubare}) and self-explanatory
notation for the renormalization constants of the Green functions with
BRS sources. The latter Green functions are themselves constrained by
Slavnov-Taylor identities relating the self energies of the scalar and
fermionic components of chiral SUSY multiplets. The respective STIs
have been shown to be valid on the regularized 2-loop level in
Ref.~\cite{DS05}, and as a result we know that the respective
renormalization constants can be expressed in terms of elementary
field renormalization constants as
\begin{subequations}
  \label{fieldrenZy}
\begin{align}
  Z_{\tilde{u}\bar\epsilon y_{u_R}} &= \sqrt{Z_{\tilde{u}_R}} /
  \sqrt{Z_{u_R}},\\
  Z_{H_2\bar\epsilon y_2} &= \sqrt{Z_{H_2}} / \sqrt{Z_{\tilde{H}_2}}.
\end{align}
\end{subequations}
Combining the previous two equations (\ref{fieldrenZy}) with the
result of the STI (\ref{resultSTI}) and assuming
Eq.\ (\ref{claimyukDef}) to hold by definition then yields the desired
result 
\begin{align}
  \label{ZHsuq}
    Z_{\tilde{H}_2q\tilde{u}} &= Z_{y_u}\sqrt{Z_{\tilde{H}_2} Z_{q_L} Z_{\tilde{u}_R}}.
\end{align}
To summarize: evaluating the functional derivative of the STI
(\ref{STIexample}) leads to the specific STI (\ref{STIqsqH}), which is
valid on the regularized 2-loop level in the gaugeless limit. As a
result we can prove the second identity of our claim
Eq.\ (\ref{claimyuksector}) in the desired order. All
these steps can be repeated for the case in which the role of $q_L$
and $u_R$ are exchanged, i.e.\ for the functional derivative
w.r.t.\ the set of fields $\tilde{q}_L^i \bar{u}_R H_2^j
\epsilon$. Without going through the details it is clear that the
result is then given by
\begin{align}
  \label{ZHsqu}
  Z_{\tilde{H}_2\tilde{q}{u}} &= Z_{y_u}\sqrt{Z_{\tilde{H}_2}
    Z_{\tilde{q}_L} Z_{u_R}} \,,
\end{align}
the third and final identity of Eq.\ (\ref{claimyuksector}).

\subsection{Other lower-order cases}
\label{sec:otherSTI}

We will now present a list of further STIs, which can be treated in an
analogous way. In all cases we have verified that the identities are
satisfied in the gaugeless limit at the 2-loop level with 1-loop
subrenormalization, and we have derived the resulting identities
between renormalization constants.
 The list is as follows:
\begin{itemize}
\item Starting from the same identity as (\ref{STIexample}),
  \begin{align}
  0&=\frac{\delta^4 S(\Gamma)}{\delta q_L^i \delta \tilde{u}_R^\dagger
    \delta H_2^j\delta \bar\epsilon}\,,
  \end{align}
  but evaluating it for the case in which the Higgs field carries no
  momentum, $p_{H}=0$ but $p_{q_L}\ne0$, leads to
  \begin{align}
  0=&
  - \Gamma_{\tilde{u}_R^\dagger\,\bar\epsilon \, y_{u_R}}
  \Gamma_{q_L^i \,H_2^j \,\bar{u}_R}
  - \Gamma_{H_2^j \,\tilde{u}_R^\dagger \,\bar\epsilon \,y_{q_L^l}}
  \Gamma_{q_L^i \, \bar{q}_L^l } +\ldots \, .
\end{align}
Here and in the subsequent equations the dots have the analogous meaning to
Eq.\ (\ref{STIqsqH}). Using self-explanatory notation, this leads to
the following identity between renormalization constants:
\begin{align}
  Z_{\tilde{u}\bar\epsilon y_{u_R}} Z_{H_2q{u}} &= Z_{H_2\tilde{u}
    \epsilon y_{q_L}} Z_{q_L} \, .
\end{align}
This determines the renormalization of the BRS transformation of the
quark field $q_L$ into an F-term given by the product
$H_2\tilde{u}_R^\dagger$:
\begin{align}
  \label{ZHsuyq}
  Z_{H_2\tilde{u}
    \epsilon y_{q_L}} &=
  Z_{y_u}\sqrt{Z_{H_2}Z_{\tilde{u}_R}/ Z_{q_L}} \, ,
\end{align}
in agreement with the symmetric renormalization transformation
(\ref{SymRenTrans}). 
\item An analogous identity with $q_L$ and $u_R$ exchanged yields
\begin{align}  
  Z_{\tilde{q}\epsilon y_{q_L}} Z_{H_2q{u}} &= Z_{H_2\tilde{q}
    \epsilon y_{u_R}} Z_{u_R} \, 
\end{align}
and determines the renormalization of the BRS transformation of the
quark field $u_R$ into the appropriate F-term:
\begin{align}
  \label{ZHsqyu}
  Z_{H_2\tilde{q}
    \epsilon y_{u_R}} &=
  Z_{y_u}\sqrt{Z_{H_2}Z_{\tilde{q}_L}/ Z_{u_R}} \, .
\end{align}
\item Taking the functional derivative w.r.t.\ the set of
  fields $\tilde{q}_L^i \tilde{u}_R^\dagger \tilde{H}_2^j
  \bar\epsilon$ and evaluating it for the case with
  $p_{\tilde{q}_L}=0$ but $p_{\tilde{H}_2}\ne0$ leads to
  \begin{align}
    0&=
    - \Gamma_{\tilde{u}_R^\dagger\,\bar\epsilon \, y_{u_R}}
  \Gamma_{\tilde{q}_L^i \,\tilde{H}_2^j \,\bar{u}_R}
  - \Gamma_{\tilde{q}_L^j \,\tilde{u}_R^\dagger \,\bar\epsilon \,y_{\tilde{H}_l^m}}
  \Gamma_{\tilde{H}_2^i \, \overline{\tilde{H}_l^m}}
  +\ldots \, .
\end{align}
  The resulting identity between renormalization constants reads
  \begin{align}
    Z_{\tilde{u}\epsilon y_{u_R}} Z_{\tilde{H}_2\tilde{q}u} &=
    Z_{\tilde{q}\tilde{u}\epsilon y_{\tilde{H}}} Z_{\tilde{H}_2}
  \end{align}
  and determines the renormalization of the BRS transformation of the
  Higgsino field $\tilde{H}_2$ into the appropriate F-term:
  \begin{align}
    \label{Zsqsuyh}
    Z_{\tilde{q}\tilde{u}
      \epsilon y_{\tilde{H}_2}} &=
    Z_{y_u}\sqrt{ Z_{\tilde{q}_L}Z_{\tilde{u}_R} / Z_{\tilde{H}_2}} \, .
  \end{align}
\item
  Taking the functional derivative w.r.t.\ the set of fields
  $\tilde{u}_R \tilde{u}_R^\dagger H_2^j \tilde{H}_2^i{}^C
  \bar\epsilon$ and evaluating it leads to
  \begin{align}
    0&= 
    - \Gamma_{H_2^j \,\tilde{u}_R^\dagger \,\bar\epsilon \,y_{q_L^l}}
    \Gamma_{\tilde{u}_R \, \tilde{H}_2^i{}^C \, \bar{q}_L^l}
    + \Gamma_{\tilde{H}_2^i{}^C \, \bar\epsilon \,
      Y_{H_2^l{}^\dagger}}
    \Gamma_{\tilde{u}_R \, \tilde{u}_R^\dagger \, H_2^j \,
      H_2^l{}^\dagger}
    +\ldots
    \, .
  \end{align}
  The resulting identity between renormalization constants reads
  \begin{align}
    Z_{H_2\tilde{u} \epsilon y_{q_L}}  Z_{\tilde{H}_2q\tilde{u}} &=
    Z_{\tilde{H}_2\epsilon Y_{H_2}} Z_{\tilde{u}\tilde{u}H_2 H_2} \, .
  \end{align}
  This identity determines the quartic interaction between two
  right-handed squarks
  $\tilde{u}_R$ and two Higgs bosons $H_2$ in terms of previously
  determined renormalization constants:
  \begin{align}
    \label{ZsusuHH}
    Z_{\tilde{u}\tilde{u}H_2 H_2} &=
    Z_{y_u}^2 Z_{H_2} Z_{\tilde{u}_R} \, .
  \end{align}
\item
  A similar identity which contains $\tilde{q}_L$ instead of
  $\tilde{u}_R$ determines the renormalization constant
  $Z_{\tilde{q}\tilde{q}H_2 H_2}$ between two left-handed squarks
  $\tilde{q}_L$ and two Higgs bosons $H_2$:
  \begin{align}
    \label{ZsqsqHH}
    Z_{\tilde{q}\tilde{q}H_2 H_2} &=
    Z_{y_u}^2 Z_{H_2} Z_{\tilde{q}_L} \, .
  \end{align}
\item Taking the functional derivative w.r.t.\ the set of
  fields $A^\mu_a A^\nu_b H_2^i \tilde{H}_2^j
  \bar\epsilon$ leads to the identity
  \begin{align}
    0 &=
    \Gamma_{ \tilde{H}_2^j \, \bar\epsilon Y_{H_2^l{}^\dagger}}
    \Gamma_{A^\mu_a \, A^\nu_b \, H_2^i H_2^l{}^\dagger}
    +\ldots \, .
  \end{align}
  This identity determines the interaction of two gluons and two Higgs
  bosons. Again, the identity is valid in DRED on the regularized
  2-loop level. This is true in particular for the $D$-dimensional
  gauge fields but also for the so-called
  $\epsilon$-scalar part of the gluons, i.e.\ the $(4-D)$ extra gluon
  components which appear on the regularized level and which behave
  like adjoint scalar fields and not like gauge fields.   Following
  e.g.\ Ref.\ \cite{DS05} we denote the $D$-dimensional gauge field
  gluons as $\hat{A}^\mu_a$ and the $\epsilon$-scalars as
  $\tilde{A}^\mu_a$. As a result of the identity,  the renormalization
  constants for both the 
  quartic $\hat{A}\hat{A}H_2H_2$ and the quartic
  $\tilde{A}\tilde{A}H_2H_2$ interactions are determined to   vanish,  
  \begin{align}
    \label{ZAAHH}
    Z_{\hat{A}\hat{A}H_2 H_2} &= 0 &
        Z_{\tilde{A}\tilde{A}H_2 H_2} &= 0\, .
  \end{align}
  The first of these equations would also follow from gauge
  invariance, but the second would not. If the renormalization
  constant $        Z_{\tilde{A}\tilde{A}H_2 H_2}$ would not vanish at
  the 2-loop level,   it could contribute to Higgs boson self
  energies at the 3-loop level.
\item
  All previous identities involve the Higgs doublet $H_2$ and/or
  right-handed up-(s)quarks. Identities involving $H_1$ and/or
  right-handed down-(s)quarks can be derived in the same way, with
  analogous results.
\end{itemize}
The previous identities are summarized in Tab.\ \ref{tabsti}. They
lead to Eqs.\ 
(\ref{ZHsuq},\ref{ZHsqu},\ref{ZHsuyq},\ref{ZHsqyu},\ref{Zsqsuyh},\ref{ZsusuHH},\ref{ZsqsqHH},\ref{ZAAHH})
and thus
determine a variety of renormalization 
constants. In all cases, the results agree with the symmetric counterterms
generated by the renormalization transformation
(\ref{SymRenTrans}). The considered cases cover all Green functions with up 
to two Higgs/Higgsino fields and up to two coloured fields, up to the 2-loop
level in the gaugeless limit. They include not only physical fields but
also $\epsilon$-scalars, sources for BRS transformations and SUSY
ghosts. In other words, the previous identities establish that 
SUSY-restoring counterterms are not required in this considered
sector. 

In addition, Refs.\ \cite{Harlander:2006xq,Harlander:2009mn} have
shown the absence of SUSY-restoring 
counterterms in the pure-SUSY-QCD sector of triple interactions
between quarks, squarks, gluons, gluinos and
$\epsilon$-scalars. Quartic interactions involving four coloured
fields are not covered by any of these analyses.

\begin{table}[h]
\begin{center}
\begin{tabular}{c| c |c}
%% \toprule [0.1em]
  STI  &  Relevant part & Expressed symmetry properties  \\
\hline
  %  \specialrule{0.1em}{0em}{0em} & &  \\[-1.0em] 
 \multirow{2}{*}{$\epsilon  \tilde q_L^{\dagger} q_L $} &
 \multirow{2}{*}{$\propto p_q^2$} & Connects the SUSY transformations
 of \\
 && squark into quark and vice versa to \\
 \multirow{2}{*}{$\epsilon \bar\epsilon \tilde q_L^{\dagger} Y_{\tilde q_L} $} &
 \multirow{2}{*}{$\propto \psl$} &  field renormalizations (from Ref.\ \cite{DS05}; \\
 && similar for $u_R$ and for Higgs/Higgsino)\\ \hline & &  \\[-1.0em] 
 \multirow{4}{*}{$q_L \tilde{u}_R^\dagger
    H_2 \bar\epsilon$} &  \multirow{4}{*}{$ p_{q_L} = 0 $} &
 Connects the Yukawa coupling of \\
 && the Higgs boson with quarks to the \\
 %% \multirow{2}{*}{$\tilde{q}_L \bar{u}_R H_2
 %%   \epsilon$}
 & %% \multirow{2}{*}{$ p_{u_R} = 0 $}
 &  Yukawa coupling of quark/squark\\
 &&  and Higgsino (similar for $\tilde{q}_L\bar{u}_R $)
 \\ \hline & &  \\[-1.0em]  
 \multirow{3}{*}{same} &  \multirow{3}{*}{$ p_{H} = 0 $} & Determines
 SUSY transformation     \\
 & & of quark into an F-term \\
 & &  involving Higgs and squark field   \\ \hline & &  \\[-1.0em] 
  \multirow{3}{*}{$\tilde{q}_L \tilde{u}_R^\dagger \tilde{H}_2
  \bar\epsilon$} &  \multirow{3}{*}{$ p_{q_L} =  0$} & Determines
 SUSY transformation    \\
 & &  of Higgsino into an F-term \\
 & &  involving two squark fields   \\
 \hline & &  \\[-1.0em] 
  \multirow{3}{*}{$\tilde{u}_R \tilde{u}_R^\dagger H_2 \tilde{H}_2{}^C
  \bar\epsilon$} &  \multirow{2}{*}{$ p = 0 $} & Determines quartic
  interactions
  \\
  %% $\tilde{q}_L \tilde{q}_L^\dagger H_2 \tilde{H}_2{}^C
  %% \bar\epsilon$
  & & between Higgs and squark fields\\
  & & (similar for $\tilde{q}_L$) \\
  \hline  & &  \\[-1.0em] 
 \multirow{3}{*}{$A^\mu A^\nu H_2 \tilde{H}_2
  \bar\epsilon$} &  \multirow{3}{*}{$ p = 0$} & Determines quartic coupling between\\
 & &  Higgs bosons and gluons or\\
 & & $\epsilon$-scalars  \\ 
%$ \epsilon \bar \epsilon  \tilde q^{\dagger} Y_{\tilde q^{\dagger}}$ & $ \epsilon \overline \psi_1 A^{\mu} h h,$&  Connects minimal coupling term to  \\
% \bottomrule [0.1em]
\end{tabular}
\end{center}
\caption{Summary of the SUSY STIs used in sec.\ \ref{sec:lowerorder}
  for the 2-loop  counterterms for subrenormalization.} \label{tabsti}
\end{table}

\section{3-loop results}
\label{sec:threeloop}

\begin{figure}[tb]
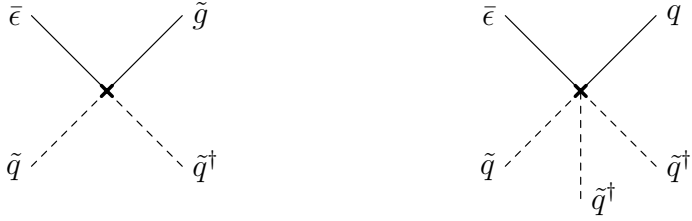

\begin{center}
\unitlength=0.9cm%
\begin{feynartspicture}(16,3.5)(3,1)
\FADiagram{}
\FAVert(10,10){1}
\FALabel(3,17)[r]{$\epsbar$\ }
\FAProp(17,17)(10,10)(0.,){/Straight}{0}
\FALabel(17,17)[l]{\ ${\glui}$}
\FAProp(3,17)(10,10)(0.,){/Straight}{0}
\FALabel(3,3)[r]{$\sq$\ }
\FAProp(17,3)(10,10)(0.,){/ScalarDash}{0}
\FALabel(17,3)[l]{\ ${\sq^\dagger}$}
\FAProp(3,3)(10,10)(0.,){/ScalarDash}{0}
\FADiagram{}
\FADiagram{}
\FAVert(10,10){1}
\FALabel(3,17)[r]{$\epsbar$\ }
\FAProp(17,17)(10,10)(0.,){/Straight}{0}
\FALabel(17,17)[l]{\ ${q}$}
\FAProp(3,17)(10,10)(0.,){/Straight}{0}
\FALabel(3,3)[r]{$\sq$\ }
\FAProp(3,3)(10,10)(0.,){/ScalarDash}{0}
\FALabel(17,3)[l]{\ ${\sq^\dagger}$}
\FAProp(17,3)(10,10)(0.,){/ScalarDash}{0}
\FALabel(10,0)[l]{\ ${\sq^\dagger}$}
\FAProp(10,0)(10,10)(0.,){/ScalarDash}{0}
\end{feynartspicture}
\end{center}
\vspace{-.5cm}
\caption{Possible Feynman rules corresponding to
  $\Delta^{\text{2L}}_{\text{extra}}$ in
  Eq.\ (\ref{DefDeltaextra}). The first Feynman rule could appear if
  a non-SUSY 2-loop counterterm to an interaction of the type
  $\ygluibar\sq^\dagger \sq$ is required, where $\ygluibar$ is the source of the
  gluino BRS transformation. The second Feynman rule could appear if a
  non-SUSY 2-loop counterterm to a 4-squark interaction is
  required. Terms in   $\Delta^{\text{2L}}_{\text{extra}}$ with less
  than three coloured fields are excluded by the discussion of
  Sec.\ \ref{sec:lowerorder}. 
}
\label{fig:FeynmanrulesDelta4scalar}
\end{figure}

In this section we determine the potential non-SUSY counterterms which
might enter the Higgs boson mass calculation at the 3-loop level in
the gaugeless limit, i.e.\ at the orders ${\cal
  O}(\alpha_{t,b}\alpha_s^2,\alpha_{t,b}^2\alpha_s,\alpha_{t,b}^3)$. Ref.~\cite{HollikDS05}
has identified the relevant STI. It 
is obtained by taking 
%The following derivative of the Slavnov-Taylor identity directly
%constrains the desired Higgs-boson four-point functions:
\begin{align}
0&=\frac{\delta^5  S(\Gamma)}{\delta\phi_a\delta\phi_b\delta\phi_c
\delta\tilde{H}_{kL}^l\delta\bar\epsilon},
\label{phi4STI1}
\end{align}
where $\phi_i$ denote any components of the MSSM Higgs bosons $H_i^j$,
$H_i^j{}^\dagger$, and $\tilde{H}_k^l$ is the Higgsino partner of
$H_k^l$. Evaluating the derivative without taking the gaugeless limit
leads to the following identity:
\begin{align}
0=&
\sum_{\phi_i}
\Gamma_{\tilde{H}_{kL}^lY_{\phi_i}\bar\epsilon}
\Gamma_{\phi_a\phi_b\phi_c\phi_i}
%+
%\Gamma_{\phi_a\phi_b\phi_c\tilde{H}_{kL}^lY_{\phi_i}\bar\epsilon_L}
%\Gamma_{\phi_i}
%\nonumber\\
%+\bigg(
%&
%\Gamma_{\phi_a\tilde{H}_{kL}^lY_{\phi_i}\bar\epsilon_L}
%\Gamma_{\phi_b\phi_c\phi_i}
%+
%\Gamma_{\phi_a\phi_b\tilde{H}_{kL}^lY_{\phi_i}\bar\epsilon_L}
%\Gamma_{\phi_c\phi_i}
%+\mbox{perm}
%\bigg)
+
\sum_{\lambda}
\Gamma_{ \phi_a\phi_b Y_{\lambda}\bar\epsilon}
\Gamma_{\tilde{H}_{kL}^l\phi_c\lambda}
+\text{perm.}+\text{fin.}
%% \nonumber\\
%% -
%% \sum_{i,j}\bigg[
%% %&
%% %\Gamma_{y_i^j\bar\epsilon_L}
%% %\Gamma_{\phi_a\phi_b\phi_c \tilde{H}_{kL}^l\overline{\tilde{H}}_i^j}
%% %+
%% %\Gamma_{y_i^{jC}\bar\epsilon_L}
%% %\Gamma_{\phi_a\phi_b\phi_c \tilde{H}_{kL}^l\overline{\tilde{H}}_i^{jC}}
%% %\nonumber\\
%% %+
%% %\bigg(&
%% %\Gamma_{\phi_a y_i^j\bar\epsilon_L}
%% %\Gamma_{\phi_b\phi_c \tilde{H}_{kL}^l\overline{\tilde{H}}_i^j}
%% %+
%% %\Gamma_{\phi_a y_i^{jC}\bar\epsilon_L}
%% %\Gamma_{\phi_b\phi_c\tilde{H}_{kL}^l\overline{\tilde{H}}_i^{jC}}
%% %+\mbox{perm}\bigg)\nonumber\\
%% %+
%% %\bigg(&
%% %\Gamma_{\phi_a\phi_b y_i^j\bar\epsilon_L}
%% %\Gamma_{\phi_c\tilde{H}_{kL}^l\overline{\tilde{H}}_i^j}
%% %+
%% %\Gamma_{\phi_a\phi_b y_i^{jC}\bar\epsilon_L}
%% %\Gamma_{\phi_c\tilde{H}_{kL}^l\overline{\tilde{H}}_i^{jC}}
%% %+\mbox{perm}\bigg)\nonumber\\
%% +
%% &
%% \Gamma_{\phi_a\phi_b\phi_c y_i^j\bar\epsilon_L}
%% \Gamma_{\tilde{H}_{kL}^l\overline{\tilde{H}}_i^j}
%% +
%% \Gamma_{\phi_a\phi_b\phi_c y_i^{jC}\bar\epsilon_L}
%% \Gamma_{\tilde{H}_{kL}^l\overline{\tilde{H}}_i^{jC}}
%% \bigg]
%% \nonumber\\
%% +&\sqrt2 P_L f_0 \Gamma_{\phi_a\phi_b\phi_c \tilde{H}_{kL}^l
%% \overline{\chi}_L^C}
%% .
\label{phi4STI}
\end{align}
Here the abbreviation ``fin.'' summarizes terms which vanish at
tree-level and which at $n$-loop order don't receive $n$-loop
counterterm contributions; ``perm'' denotes terms 
corresponding to all possible permutations of $\phi_{a,b,c}$. The sums
run over all Higgs field components $\phi_i$ and all electroweak
gauginos $\lambda$.
This identity describes the fundamental
supersymmetry relation between the quartic Higgs-boson self-coupling
and the electroweak gauge couplings, which is behind all Higgs-boson mass
predictions in the MSSM. The gauge couplings are reflected in the
second term of 
Eq.~(\ref{phi4STI}), which corresponds to the SUSY transformation of
electroweak gauginos into the $D$-terms, which in turn contain
products of gauge couplings and Higgs fields.

Although this identity is rather involved, it determines unambiguously
the desired counterterm. From the arguments given in Ref.~\cite{HollikDS05} we
obtain immediately the following
implication: {\em ``If Eq.~(\ref{phi4STI}) is valid on the regularized
  $n$-loop level in the gaugeless limit (with $(n-1)$-loop
  subrenormalization), then the potential non-SUSY counterterms to the
  Higgs sector are zero.''} 

Hence we will now study the identity at
the 3-loop level with 2-loop subrenormalization.
Using again the quantum action principle, we can write the potential
breaking of the identity (\ref{phi4STI}) at this level as
\begin{align}
\left([\Delta^{\le\text{2L}}]\cdot
\Gamma^{\rm DRED}\right)_{\phi_a\phi_b\phi_c\tilde{H}_{kL}^l\bar\epsilon}
\label{breaking3loopSTI}
\end{align}
where now $\Delta^{\le\text{2L}}\equiv
S(\Gamma_{\text{cl+ct}}^{\le\text{2L}})$ and where
$\Gamma_{\text{cl+ct}}^{\le\text{2L}}$ is the 
2-loop bare action. We will evaluate the potential breaking at
zero external momenta. This is sufficient since all Green functions appearing in the
STI (\ref{phi4STI}) are of dimension 4, and correspondingly the
quartic Higgs counterterm to be determined is momentum-independent. 

First we need to clarify the structure of the insertion
$\Delta^{\le\text{2L}}$ and its Feynman rules. As discussed in the previous section the
2-loop counterterm structure is given essentially by the usual
SUSY-preserving renormalization transformation, but the given proof
does not exclude non-SUSY counterterms to quartic interactions of four
coloured scalars (squarks or $\epsilon$-scalars). Hence we can write
the insertion as 
\begin{align}
\Delta^{\le\text{2L}}\equiv
S(\Gamma_{\text{cl+ct}}^{\le\text{2L}}) &=
\Delta_{\text{ren.transf.}}+\Delta^{\text{2L}}_{\text{extra}}
\,,
\label{DefDeltaextra}
\end{align}
where the first term is obtained from the result given in Ref.~\cite{DS05}
by the symmetric renormalization transformation
(\ref{SymRenTrans}). The second term might 
be present in 
the case that non-SUSY 2-loop counterterms are
indeed necessary; non-SUSY 2-loop counterterms are constrained by the
discussion of the previous section. Hence
$\Delta^{\text{2L}}_{\text{extra}}$ would contain at least three
powers of coloured 
fields and lead to Feynman rules like the ones shown in
Fig.~\ref{fig:FeynmanrulesDelta4scalar}. It could contribute to
Eq.~(\ref{breaking3loopSTI}) earliest if inserted into 2-loop diagrams 
but is itself of 2-loop order. Hence for the purpose of the 3-loop
evaluation of Eq.~(\ref{breaking3loopSTI}) we can ignore
$\Delta^{\text{2L}}_{\text{extra}}$. The relevant Feynman rules for
$\Delta^{\le\text{2L}}$ are then the ones given in Ref.~\cite{DS05}, up to the usual
renormalization transformation.

\begin{figure}[tb]
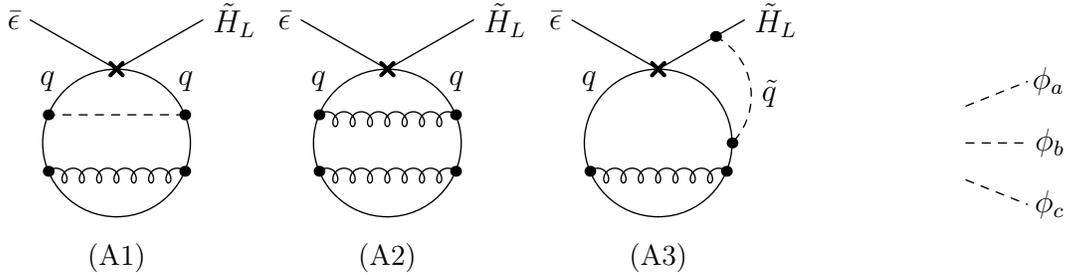

\begin{center}
\unitlength=0.9cm%
\begin{feynartspicture}(16,4)(4,1)
\FADiagram{(A1)}
\FAVert(10,16){1}
\FALabel(3,20)[r]{$\epsbar$\ }
\FAProp(17,20)(10,16)(0.,){/Straight}{0}
\FALabel(17,20)[l]{\ ${\tilde{H}_L}$}
\FAProp(3,20)(10,16)(0.,){/Straight}{0}
\FALabel(5,15)[r]{$q$}
\FALabel(15,15)[l]{$q$}
%\FALabel(16,7)[l]{$q$}
%\FALabel(4,7)[r]{$\tilde{H}$}
%\FALabel(10,4.5)[b]{$q$}
%\FALabel(10,10.5)[b]{$\tilde{q}$}
%\FAProp(4,10)(16,10)(0.,){/ScalarDash}{0}
\FAProp(10,16)(10,4)(1.,){/Straight}{0}
\FAProp(10,16)(10,4)(-1.,){/Straight}{0}
\FAVert(15.54,12.29){0}
\FAVert(4.45,12.29){0}
\FAVert(15.54,7.7){0}
\FAVert(4.45,7.7){0}
\FAProp(15.54,12.29)(4.45,12.29)(0.,){/ScalarDash}{0}
\FAProp(15.54,7.7)(4.45,7.70)(0.,){/Cycles}{0}
\FADiagram{(A2)}
\FAVert(10,16){1}
\FALabel(3,20)[r]{$\epsbar$\ }
\FAProp(17,20)(10,16)(0.,){/Straight}{0}
\FALabel(17,20)[l]{\ ${\tilde{H}_L}$}
\FAProp(3,20)(10,16)(0.,){/Straight}{0}
\FALabel(5,15)[r]{$q$}
\FALabel(15,15)[l]{$q$}
%\FALabel(16,7)[l]{$q$}
%\FALabel(4,7)[r]{$\tilde{H}$}
%\FALabel(10,4.5)[b]{$q$}
%\FALabel(10,10.5)[b]{$\tilde{q}$}
%\FAProp(4,10)(16,10)(0.,){/ScalarDash}{0}
\FAProp(10,16)(10,4)(1.,){/Straight}{0}
\FAProp(10,16)(10,4)(-1.,){/Straight}{0}
\FAVert(15.54,12.29){0}
\FAVert(4.45,12.29){0}
\FAVert(15.54,7.7){0}
\FAVert(4.45,7.7){0}
\FAProp(15.54,12.29)(4.45,12.29)(0.,){/Cycles}{0}
\FAProp(15.54,7.7)(4.45,7.70)(0.,){/Cycles}{0}
\FADiagram{(A3)}
\FAVert(10,16){1}
\FALabel(3,20)[r]{$\epsbar$\ }
\FAProp(17,20)(10,16)(0.,){/Straight}{0}
\FALabel(17,20)[l]{\ ${\tilde{H}_L}$}
\FAProp(3,20)(10,16)(0.,){/Straight}{0}
\FALabel(5,15)[r]{$q$}
%\FALabel(15,15)[l]{$\tilde{H},\glui$}
\FAProp(10,16)(10,4)(1.,){/Straight}{0}
\FAProp(10,16)(10,4)(-1.,){/Straight}{0}
\FAVert(15.54,7.7){0}
\FAVert(4.45,7.7){0}
\FAProp(15.54,7.7)(4.45,7.70)(0.,){/Cycles}{0}
\FAProp(14.66,18.66)(16,10)(-.5,){/ScalarDash}{0}
\FAVert(14.66,18.66){0}
\FAVert(16,10){0}
\FALabel(18,14)[l]{\,$\tilde{q}$}
\FADiagram{}
\FAProp(13,7)(18,5)(0.,){/ScalarDash}{0}
\FAProp(13,10)(18,10)(0.,){/ScalarDash}{0}
\FAProp(13,13)(18,15)(0.,){/ScalarDash}{0}
\FALabel(21,5)[r]{\ $\phi_c$}
\FALabel(21,10)[r]{\ $\phi_b$}
\FALabel(21,15)[r]{\ $\phi_a$}
\end{feynartspicture}
\end{center}
\vspace{-.5cm}
\caption{First set of diagrams contributing the potential breaking of the
  Slavnov-Taylor identity Eq.~(\ref{breaking3loopSTI}) at the 3-loop
  level in the gaugeless limit. These diagrams contain one fermion
  loop, connected by up to one boson line to the external fermion
  line. The insertion of the operator 
  $\Delta_{\text{ren.transf.}}$ is marked by a cross. Quarks, gluinos
  and Higgsinos are denoted by solid lines, gluons by circles lines;
  Higgs and squark lines are dashed.
  The lines corresponding to the external Higgs bosons $\phi_{a,b,c}$
  can be attached in all possible ways.
 % In the text, the $\gamma$-string attached to the second
 % external fermion line is denoted as $A$, the $\gamma$-string
 % attached to the closed fermion loop as $B$.
}
\label{fig:diagramsSTI1}
\end{figure}

\begin{figure}[tb]
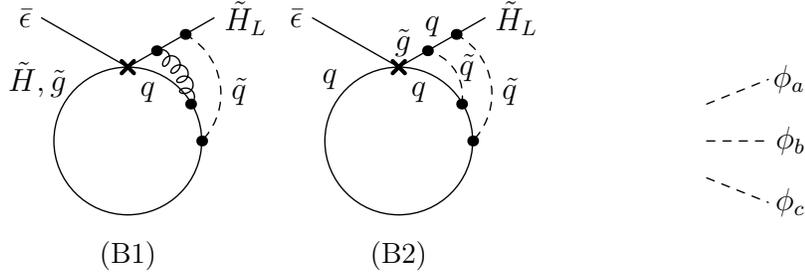

\begin{center}
\unitlength=0.9cm%
\begin{feynartspicture}(16,4)(3,1)
\FADiagram{(B1)}
\FAVert(10,16){1}
\FALabel(3,20)[r]{$\epsbar$\ }
\FAProp(17,20)(10,16)(0.,){/Straight}{0}
\FALabel(17,20)[l]{\ ${\tilde{H}_L}$}
\FAProp(3,20)(10,16)(0.,){/Straight}{0}
\FALabel(5,15)[r]{$\tilde{H},\glui$}
\FALabel(11,14)[l]{$q$}
\FAProp(10,16)(10,4)(1.,){/Straight}{0}
\FAProp(10,16)(10,4)(-1.,){/Straight}{0}
\FAVert(15.1,13){0}
\FAVert(12.33,17.33){0}
\FAProp(15.1,13)(12.33,17.33)(0.4,){/Cycles}{0}
\FAProp(14.66,18.66)(16,10)(-.5,){/ScalarDash}{0}
\FAVert(14.66,18.66){0}
\FAVert(16,10){0}
\FALabel(18,14)[l]{\,$\tilde{q}$}
\FADiagram{(B2)}
\FAVert(10,16){1}
\FALabel(3,20)[r]{$\epsbar$\ }
\FAProp(17,20)(10,16)(0.,){/Straight}{0}
\FALabel(17,20)[l]{\ ${\tilde{H}_L}$}
\FAProp(3,20)(10,16)(0.,){/Straight}{0}
\FALabel(11,18)[r]{$\glui$}
\FALabel(13.33,19.33)[r]{$q$}
\FALabel(5,15)[r]{$q$}
\FALabel(11,14)[l]{$q$}
\FAProp(10,16)(10,4)(1.,){/Straight}{0}
\FAProp(10,16)(10,4)(-1.,){/Straight}{0}
\FAVert(15.1,13){0}
\FAVert(12.33,17.33){0}
\FAProp(15.1,13)(12.33,17.33)(0.4,){/ScalarDash}{0}
\FAProp(14.66,18.66)(16,10)(-.5,){/ScalarDash}{0}
\FAVert(14.66,18.66){0}
\FAVert(16,10){0}
\FALabel(14.5,16)[l]{\,$\tilde{q}$}
\FALabel(18,14)[l]{\,$\tilde{q}$}
\FADiagram{}
\FAProp(13,7)(18,5)(0.,){/ScalarDash}{0}
\FAProp(13,10)(18,10)(0.,){/ScalarDash}{0}
\FAProp(13,13)(18,15)(0.,){/ScalarDash}{0}
\FALabel(21,5)[r]{\ $\phi_c$}
\FALabel(21,10)[r]{\ $\phi_b$}
\FALabel(21,15)[r]{\ $\phi_a$}
\end{feynartspicture}
\end{center}
\vspace{-.5cm}
\caption{Second set of diagrams contributing the potential breaking of the
  Slavnov-Taylor identity Eq.~(\ref{breaking3loopSTI}) at the 3-loop
  level in the gaugeless limit. These diagrams contain one fermion
  loop, connected by two boson lines to the external fermion
  line. The notation is as in Fig.~\ref{fig:diagramsSTI1}.
}
\label{fig:diagramsSTI2}
\end{figure}

\begin{figure}[tb]
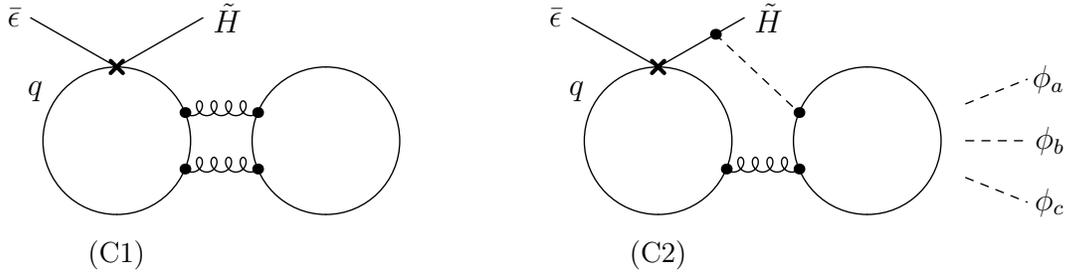

\begin{center}
\unitlength=0.9cm%
\begin{feynartspicture}(16,4)(4,1)
\FADiagram{(C1)}
\FAVert(10,16){1}
\FALabel(3,20)[r]{$\epsbar$\ }
\FAProp(17,20)(10,16)(0.,){/Straight}{0}
\FALabel(17,20)[l]{\ ${\tilde{H}}$}
\FAProp(3,20)(10,16)(0.,){/Straight}{0}
\FALabel(4,14)[r]{$q$}
\FAProp(10,16)(10,4)(1.,){/Straight}{0}
\FAProp(10,16)(10,4)(-1.,){/Straight}{0}
\FAProp(27,16)(27,4)(1.,){/Straight}{0}
\FAProp(27,16)(27,4)(-1.,){/Straight}{0}
\FAVert(15.54,12.29){0}
\FAVert(21.45,12.29){0}
\FAVert(15.54,7.7){0}
\FAVert(21.45,7.7){0}
\FAProp(15.54,12.29)(21.45,12.29)(0.,){/Cycles}{0}
\FAProp(15.54,7.7)(21.45,7.70)(0.,){/Cycles}{0}
\FADiagram{}
\FADiagram{(C2)}
\FAVert(10,16){1}
\FALabel(3,20)[r]{$\epsbar$\ }
\FAProp(17,20)(10,16)(0.,){/Straight}{0}
\FALabel(17,20)[l]{\ ${\tilde{H}}$}
\FAProp(3,20)(10,16)(0.,){/Straight}{0}
\FALabel(4,14)[r]{$q$}
\FAProp(10,16)(10,4)(1.,){/Straight}{0}
\FAProp(10,16)(10,4)(-1.,){/Straight}{0}
\FAProp(27,16)(27,4)(1.,){/Straight}{0}
\FAProp(27,16)(27,4)(-1.,){/Straight}{0}
\FAVert(14.66,18.66){0}
\FAVert(21.45,12.29){0}
\FAVert(15.54,7.7){0}
\FAVert(21.45,7.7){0}
\FAProp(14.66,18.66)(21.45,12.29)(0.,){/ScalarDash}{0}
\FAProp(15.54,7.7)(21.45,7.70)(0.,){/Cycles}{0}
\FADiagram{}
\FAProp(13,7)(18,5)(0.,){/ScalarDash}{0}
\FAProp(13,10)(18,10)(0.,){/ScalarDash}{0}
\FAProp(13,13)(18,15)(0.,){/ScalarDash}{0}
\FALabel(21,5)[r]{\ $\phi_c$}
\FALabel(21,10)[r]{\ $\phi_b$}
\FALabel(21,15)[r]{\ $\phi_a$}
\end{feynartspicture}
\end{center}
\vspace{-.5cm}
\caption{Third set of diagrams contributing the potential breaking of the
  Slavnov-Taylor identity Eq.~(\ref{breaking3loopSTI}) at the 3-loop
  level in the gaugeless limit. These diagrams contain two fermion
  loops. The notation is as in Fig.~\ref{fig:diagramsSTI1}.
}
\label{fig:diagramsSTI3}
\end{figure}

Figs.~\ref{fig:diagramsSTI1}, \ref{fig:diagramsSTI2},
\ref{fig:diagramsSTI3} show representative 3-loop diagrams 
contributing to the potential breaking of the STI, i.e.\ to the Green
function in Eq.~(\ref{breaking3loopSTI}). Similar to the 2-loop case
of Fig.~\ref{fig:breakingexample}, each diagram contains one open
fermion line and one fermion loop connected to the insertion
$\Delta_{\text{ren.transf.}}$. We will again refer to the $\gamma$-string of the open
fermion line as $A$ and the $\gamma$-string of the fermion loop as
$B$.
The diagrams can be 
classified into three classes, according to the number of boson lines
connecting $A$ and $B$ and according to whether there is a second
fermion loop. We now discuss each class in turn; a summary can be
found in Tab.~\ref{tab:summarySTIclasses}.


The diagrams shown in Fig.~\ref{fig:diagramsSTI1} contain up to one
boson line connecting $A$ and $B$ and only a single fermion loop. All
such diagrams are of Topology (c) (see Tab.~\ref{tab:topologies}). The
fermion attached to the insertion in line $A$ is either a Higgsino or
quark; the connecting boson line can only be a squark line. After
integrating over all loop momenta in diagrams (A1) and (A2) and over
the fermion and gluon loop momenta in diagram (A3), the resulting
$\gamma$-string $B$ can at most depend on the single covariant $\ksl$,
where $k$ is the remaining loop momentum. As summarized in
Tab.~\ref{tab:topologies}, the criteria given in Ref.\ \cite{DS05}
allow up to 3 $\gamma$ matrices. Hence all diagrams of this class are
shown to vanish.

The diagrams shown in Fig.~\ref{fig:diagramsSTI2} involve two boson
lines connecting the external fermion line and the fermion loop. One
of the two boson lines can be a gluon --- in this case the other one
must be a squark, and the diagram must be of
topology (c), as in diagram (B1). If both connecting bosons are
squarks, the diagram can be of topology (b), as in diagram (B2). After
integrating over the fermion loop momentum, the $\gamma$-string $B$ of
these diagrams can depend on the two remaining loop momenta $k_1$,
$k_2$. Thus in diagrams like (B1) the $\gamma$-string $B$ can be
reduced to at most three $\gamma$-matrices $\ksl_1\ksl_2\gamma^\mu$,
where $\gamma^\mu$ corresponds to the gluon vertex. In diagrams like
(B2), the $\gamma$-string $B$ can be reduced to at most two
$\gamma$-matrices $\ksl_1\ksl_2$.  Hence again in all cases the
criteria of Tab.~\ref{tab:topologies} show that the diagrams vanish.

The diagrams shown in Fig.~\ref{fig:diagramsSTI3} contain a second
fermion loop. The diagrams are of topology (c).
After integrating over the two fermion-loop momenta, the
$\gamma$-string $B$ can be reduced to at most three covariants
$\ksl\gamma^\mu\gamma^\nu$, where $k$ is the remaining loop momentum
and where $\gamma^{\mu,\nu}$ correspond to gluon vertices. Hence again
the criteria for the diagrams to vanish are satisfied.

In summary (see Tab.~\ref{tab:summarySTIclasses}), we have shown all diagrams contributing to potential
breaking (\ref{breaking3loopSTI}) to vanish. Using this result in the
Slavnov-Taylor identity (\ref{phi4STI}) proves that
the Higgs potential counterterm vanishes,
\begin{align}
  V_{\text{quartic, ct}} &= 0
\end{align}
at the 3-loop level in the gaugeless limit. This corresponds to the
result of the symmetric renormalization transformation --- hence DRED
preserves SUSY manifestly at this level and symmetric counterterms are
sufficient.

\begin{table}
\centerline{
  \begin{tabular}{c|c|c|c|c|c}
    Class & $\#$ fermion- & $\#$ connecting & Topology &
    max. & allowed $\gamma$'s\\
         &loops &  bosons &  &appearing $\gamma$'s &\\
    \hline
    A1, A2& 1 & 0 & (c) & 0 & 3 \\
    A3    & 1 & 1 & (c) & 1 & 3 \\
    B1    & 1 & 2 & (c) & 3 & 3 \\
    B2    & 1 & 2 & (a,b,c) & 2 & 2\\
    C1    & 2 & 0 & (c) & 3 & 3 \\
    C2    & 2 & 1 & (c) & 3 & 3 
  \end{tabular}
}
\caption{\label{tab:summarySTIclasses}
  Summary of the properties of the classes of diagrams contributing to
  the breaking of the STI (\ref{breaking3loopSTI}). Representative
  diagrams can be found in Figs.~\ref{fig:diagramsSTI1},
  \ref{fig:diagramsSTI2}, \ref{fig:diagramsSTI3}. ``max.~appearing
  $\gamma$'s'' denotes the maximum number of different
  $\gamma$-matrices in the actual diagrams, see text. The column
  ``allowed $\gamma$'s'' is copied from the last column of
  Tab.~\ref{tab:topologies}. }
\end{table}

\section{Conclusions}

We have investigated DRED for Higgs boson mass calculations in the
MSSM. We have verified that DRED is consistent with SUSY at the 3-loop
level in the gaugeless limit. As a result, the usual symmetric
counterterms, generated by a symmetric renormalization transformation,
are sufficient. Concrete calculations of the MSSM Higgs boson mass at
the 3-loop order ${\cal O}(\alpha_{t,b}\alpha_s^2,
\alpha_{t,b}^2\alpha_s, \alpha_{t,b}^3)$ such as the existing
calculation of the order ${\cal O}(\alpha_{t}\alpha_s^2)$
\cite{HarlanderKantHiggs,Harlander:2017kuc} and
future extensions do not need SUSY-restoring counterterms.

The result has been obtained by extending the analysis of
Ref.\ \cite{HollikDS05} to the 3-loop level. Slavnov-Taylor identities
describing suitable SUSY relations were identified and shown to hold
at the regularized level. In this way the potential SUSY-restoring
counterterms are shown to vanish. The result includes not only the
genuine 3-loop counterterms in the Higgs sector but also --- combined
with Refs.\ \cite{Harlander:2006xq,Harlander:2009mn} --- all necessary
2-loop counterterms required for subrenormalization.
The increased complexity of the 3-loop 
case required several additional technical steps compared to the
2-loop case. Counterterms required for
subrenormalization were analyzed and the potential appearance of
non-SUSY 2-loop counterterms in the quantum
action principle was characterized. The Slavnov-Taylor identities were
evaluated in
specific limits for external momenta which are sensitive to the
desired counterterms and which enabled the application of the quantum
action principle.

\section*{Acknowledgements} DS  thanks  S.~Borowka, C.~Gnendiger,
R.~Harlander, W.~Hollik, L.~Mihaila, S.~Pa\ss ehr, A.~Signer and M.~Steinhauser for useful
discussions, and DS gratefully acknowledges hospitality and
financial support from
the University of Zurich and Paul Scherrer Institut during an early phase of this work.

\begin{flushleft}

\end{flushleft}


\begin{thebibliography}{99} 

\bibitem{LHCHiggs}%Aad:2012tfa
    G.~Aad et al.,
    %  title          = "{Observation of a new particle in the search for the
    %                     Standard Model Higgs boson with the ATLAS detector at
    %                     the LHC}",
    ATLAS collaboration,
    Phys.~Lett.~{\bf B 176} (2012) 1-29.
   %%CITATION = ARXIV:1207.7214;%%
%
%\bibitem{Chatrchyan:2012xdj}
S.~Chatrchyan et al., CMS collaboration,
%      title          = "{Observation of a new boson at a mass of 125 GeV with
%                         the CMS experiment at the LHC}",
Phys.~Lett.\ {\bf B716} (2012) 30-61.
%%CITATION = ARXIV:1207.7235;%%


%\cite{Draper:2016pys}
\bibitem{Draper:2016pys}
  P.~Draper and H.~Rzehak,
  %``A Review of Higgs Mass Calculations in Supersymmetric Models,''
  Phys.\ Rept.\  {\bf 619} (2016) 1
%  doi:10.1016/j.physrep.2016.01.001
  [arXiv:1601.01890 [hep-ph]].
  %%CITATION = doi:10.1016/j.physrep.2016.01.001;%%
  %27 citations counted in INSPIRE as of 28 Mar 2018

  
%\cite{Harlander:2008ju}
\bibitem{HarlanderKantHiggs}
%\bibitem{Harlander:2008ju}
  R.~V.~Harlander, P.~Kant, L.~Mihaila and M.~Steinhauser,
  %``Higgs boson mass in supersymmetry to three loops,''
  Phys.\ Rev.\ Lett.\  {\bf 100} (2008) 191602
   [Phys.\ Rev.\ Lett.\  {\bf 101} (2008) 039901]
 % doi:10.1103/PhysRevLett.101.039901, 10.1103/PhysRevLett.100.191602
  [arXiv:0803.0672 [hep-ph]];
  %%CITATION = doi:10.1103/PhysRevLett.101.039901, 10.1103/PhysRevLett.100.191602;%%
  %152 citations counted in INSPIRE as of 28 Mar 2018
%
%\cite{Kant:2010tf}
%\bibitem{Kant:2010tf}
%  P.~Kant, R.~V.~Harlander, L.~Mihaila and M.~Steinhauser,
  %``Light MSSM Higgs boson mass to three-loop accuracy,''
  JHEP {\bf 1008} (2010) 104
%  doi:10.1007/JHEP08(2010)104
  [arXiv:1005.5709 [hep-ph]].
  %%CITATION = doi:10.1007/JHEP08(2010)104;%%
  %154 citations counted in INSPIRE as of 28 Mar 2018

  %\cite{Kunz:2014gya}
\bibitem{Kunz:2014gya}
  D.~Kunz, L.~Mihaila and N.~Zerf,
  %``$\mathcal O(\alpha_S^2)$ corrections to the running top-Yukawa coupling and the mass of the lightest Higgs boson in the MSSM,''
  JHEP {\bf 1412} (2014) 136
%  doi:10.1007/JHEP12(2014)136
  [arXiv:1409.2297 [hep-ph]].
  %%CITATION = doi:10.1007/JHEP12(2014)136;%%
  %5 citations counted in INSPIRE as of 28 Mar 2018


%\cite{Harlander:2017kuc}
\bibitem{Harlander:2017kuc}
  R.~V.~Harlander, J.~Klappert and A.~Voigt,
  %``Higgs mass prediction in the MSSM at three-loop level in a pure $\overline{{\text {DR}}}$ context,''
  Eur.\ Phys.\ J.\ C {\bf 77} (2017) no.12,  814
%  doi:10.1140/epjc/s10052-017-5368-6
  [arXiv:1708.05720 [hep-ph]].
  %%CITATION = doi:10.1140/epjc/s10052-017-5368-6;%%
  %9 citations counted in INSPIRE as of 28 Mar 2018

%\cite{Martin:2007pg}
\bibitem{Martin:2007pg}
  S.~P.~Martin,
  %``Three-loop corrections to the lightest Higgs scalar boson mass in supersymmetry,''
  Phys.\ Rev.\ D {\bf 75} (2007) 055005
%  doi:10.1103/PhysRevD.75.055005
  [hep-ph/0701051].
  %%CITATION = doi:10.1103/PhysRevD.75.055005;%%
  %147 citations counted in INSPIRE as of 28 Mar 2018



%\cite{Degrassi:2014pfa}
\bibitem{Degrassi:2014pfa}
  G.~Degrassi, S.~Di Vita and P.~Slavich,
  %``Two-loop QCD corrections to the MSSM Higgs masses beyond the effective-potential approximation,''
  Eur.\ Phys.\ J.\ C {\bf 75} (2015) no.2,  61
%  doi:10.1140/epjc/s10052-015-3280-5
  [arXiv:1410.3432 [hep-ph]].
  %%CITATION = doi:10.1140/epjc/s10052-015-3280-5;%%
  %42 citations counted in INSPIRE as of 28 Mar 2018

%\cite{Borowka:2014wla}
\bibitem{Borowka:2014wlaetal}
  S.~Borowka, T.~Hahn, S.~Heinemeyer, G.~Heinrich and W.~Hollik,
  %``Momentum-dependent two-loop QCD corrections to the neutral Higgs-boson masses in the MSSM,''
  Eur.\ Phys.\ J.\ C {\bf 74} (2014) no.8,  2994
%  doi:10.1140/epjc/s10052-014-2994-0
  [arXiv:1404.7074 [hep-ph]];
  %%CITATION = doi:10.1140/epjc/s10052-014-2994-0;%%
  %51 citations counted in INSPIRE as of 28 Mar 2018
%
  %\cite{Borowka:2015ura}
%\bibitem{Borowka:2015ura}
%  S.~Borowka, T.~Hahn, S.~Heinemeyer, G.~Heinrich and W.~Hollik,
  %``Renormalization scheme dependence of the two-loop QCD corrections to the neutral Higgs-boson masses in the MSSM,''
  Eur.\ Phys.\ J.\ C {\bf 75} (2015) no.9,  424
%  doi:10.1140/epjc/s10052-015-3648-6
  [arXiv:1505.03133 [hep-ph]].
  %%CITATION = doi:10.1140/epjc/s10052-015-3648-6;%%
  %25 citations counted in INSPIRE as of 28 Mar 2018

  %\cite{Borowka:2018anu}
\bibitem{Borowka:2018anu}
  S.~Borowka, S.~Paßehr and G.~Weiglein,
  %``Complete two-loop QCD contributions to the lightest Higgs-boson mass in the MSSM with complex parameters,''
  arXiv:1802.09886 [hep-ph].
  %%CITATION = ARXIV:1802.09886;%%

  
\bibitem{eftresults}
  %\cite{Vega:2015fna}
%\bibitem{Vega:2015fna}
  J.~Pardo Vega and G.~Villadoro,
  %``SusyHD: Higgs mass Determination in Supersymmetry,''
  JHEP {\bf 1507} (2015) 159
%  doi:10.1007/JHEP07(2015)159
  [arXiv:1504.05200 [hep-ph]].
  %%CITATION = doi:10.1007/JHEP07(2015)159;%%
  %99 citations counted in INSPIRE as of 28 Mar 2018
%
%
%
%\cite{Draper:2013oza}
%\bibitem{Draper:2013oza}
  P.~Draper, G.~Lee and C.~E.~M.~Wagner,
  %``Precise estimates of the Higgs mass in heavy supersymmetry,''
  Phys.\ Rev.\ D {\bf 89} (2014) no.5,  055023
%  doi:10.1103/PhysRevD.89.055023
  [arXiv:1312.5743 [hep-ph]].
  %%CITATION = doi:10.1103/PhysRevD.89.055023;%%
  %88 citations counted in INSPIRE as of 28 Mar 2018
%
%\cite{Bagnaschi:2014rsa}
%\bibitem{Bagnaschi:2014rsa}
  E.~Bagnaschi, G.~F.~Giudice, P.~Slavich and A.~Strumia,
  %``Higgs Mass and Unnatural Supersymmetry,''
  JHEP {\bf 1409} (2014) 092
%  doi:10.1007/JHEP09(2014)092
  [arXiv:1407.4081 [hep-ph]].
  %%CITATION = doi:10.1007/JHEP09(2014)092;%%
  %97 citations counted in INSPIRE as of 28 Mar 2018
%
%
  %\cite{Lee:2015uza}
%\bibitem{Lee:2015uza}
  G.~Lee and C.~E.~M.~Wagner,
  %``Higgs bosons in heavy supersymmetry with an intermediate m$_A$,''
  Phys.\ Rev.\ D {\bf 92} (2015) no.7,  075032
%  doi:10.1103/PhysRevD.92.075032
  [arXiv:1508.00576 [hep-ph]].
  %%CITATION = doi:10.1103/PhysRevD.92.075032;%%
  %43 citations counted in INSPIRE as of 28 Mar 2018
  
%\cite{Bagnaschi:2017xid}
\bibitem{Bagnaschi:2017xid}
  E.~Bagnaschi, J.~Pardo Vega and P.~Slavich,
  %``Improved determination of the Higgs mass in the MSSM with heavy superpartners,''
  Eur.\ Phys.\ J.\ C {\bf 77} (2017) no.5,  334
%  doi:10.1140/epjc/s10052-017-4885-7
  [arXiv:1703.08166 [hep-ph]].
  %%CITATION = doi:10.1140/epjc/s10052-017-4885-7;%%
  %8 citations counted in INSPIRE as of 28 Mar 2018

\bibitem{hybridFH}
%\bibitem{Hahn:2013ria}
  T.~Hahn, S.~Heinemeyer, W.~Hollik, H.~Rzehak
                        and G.~Weiglein,
  %High-Precision Predictions for the Light CP-Even Higgs Boson Mass
                        %of the Minimal Supersymmetric Standard Model
                        Phys.\ Rev.\ Lett.\ {\bf 112} (2014) no.14, 141801
                        [arXiv:1312.4937 [hep-ph]].
                       %%CITATION = ARXIV:1312.4937;%%
%
%  
%\cite{Bahl:2016brp}
%\bibitem{Bahl:2016brp}
  H.~Bahl and W.~Hollik,
  %``Precise prediction for the light MSSM Higgs boson mass combining effective field theory and fixed-order calculations,''
  Eur.\ Phys.\ J.\ C {\bf 76} (2016) no.9,  499
%  doi:10.1140/epjc/s10052-016-4354-8
  [arXiv:1608.01880 [hep-ph]].
  %%CITATION = doi:10.1140/epjc/s10052-016-4354-8;%%
  %45 citations counted in INSPIRE as of 28 Mar 2018
%
%\cite{Bahl:2017aev}
%\bibitem{Bahl:2017aev}
  H.~Bahl, S.~Heinemeyer, W.~Hollik and G.~Weiglein,
  %``Reconciling EFT and hybrid calculations of the light MSSM Higgs-boson mass,''
  Eur.\ Phys.\ J.\ C {\bf 78} (2018) no.1,  57
%  doi:10.1140/epjc/s10052-018-5544-3
  [arXiv:1706.00346 [hep-ph]].
  %%CITATION = doi:10.1140/epjc/s10052-018-5544-3;%%
  %18 citations counted in INSPIRE as of 28 Mar 2018

  %\cite{Athron:2016fuq}
\bibitem{Athron:2016fuq}
  P.~Athron, J.~h.~Park, T.~Steudtner, D.~Stöckinger and A.~Voigt,
  %``Precise Higgs mass calculations in (non-)minimal supersymmetry at both high and low scales,''
  JHEP {\bf 1701} (2017) 079
%  doi:10.1007/JHEP01(2017)079
  [arXiv:1609.00371 [hep-ph]].
  %%CITATION = doi:10.1007/JHEP01(2017)079;%%
  %35 citations counted in INSPIRE as of 28 Mar 2018

  
%% \bibitem{HV} G.~'t Hooft and M.~Veltman,
%%                {\em Nucl. Phys.} {\bf B 44} (1972) 189.
%%                %%CITATION = NUPHA,B44,189;%%

\bibitem{Siegel79} W.~Siegel,
                       {\em Phys. Lett.} {\bf B 84} (1979) 193.
                       %%CITATION = PHLTA,B84,193;%%

\bibitem{CJN80}
D.~M.~Capper, D.~R.~T.~Jones and P.~van Nieuwenhuizen,
%``Regularization By Dimensional Reduction Of Supersymmetric And
%Nonsupersymmetric Gauge Theories,''
{\em Nucl.\ Phys.}{\bf\ B} {\bf 167} (1980) 479.
%%CITATION = NUPHA,B167,479;%%


%\cite{Gnendiger:2017pys}
\bibitem{Gnendiger:2017pys}
  C.~Gnendiger {\it et al.},
  %``To ${d}$, or not to ${d}$: recent developments and comparisons of regularization schemes,''
  Eur.\ Phys.\ J.\ C {\bf 77} (2017) no.7,  471
%  doi:10.1140/epjc/s10052-017-5023-2
  [arXiv:1705.01827 [hep-ph]].
  %%CITATION = doi:10.1140/epjc/s10052-017-5023-2;%%
  %15 citations counted in INSPIRE as of 28 Mar 2018


\bibitem{HollikDS05}
  W.~Hollik and D.~St\"ockinger,
  %``MSSM Higgs-boson mass predictions and two-loop non-supersymmetric
  %counterterms,''
  Phys.\ Lett.\  B {\bf 634}, 63 (2006).
%  [arXiv:hep-ph/0509298].
  %%CITATION = PHLTA,B634,63;%%

 
\bibitem{MartinVaughn}
S.~P.~Martin and M.~T.~Vaughn,
%``Regularization dependence of running couplings in softly broken
%supersymmetry,''
{\em Phys.\ Lett.}{\bf\ B} {\bf 318} (1993) 331.
%[arXiv:hep-ph/9308222].

%\cite{Mihaila:2009bn}
\bibitem{Mihaila:2009bn}
  L.~Mihaila,
  %``Two-loop parameter relations between dimensional regularization and dimensional reduction applied to SUSY-QCD,''
  Phys.\ Lett.\ B {\bf 681} (2009) 52
%  doi:10.1016/j.physletb.2009.09.058
  [arXiv:0908.3403 [hep-ph]].
  %%CITATION = doi:10.1016/j.physletb.2009.09.058;%%
  %17 citations counted in INSPIRE as of 28 Mar 2018

%\cite{Stockinger:2011gp}
\bibitem{Stockinger:2011gp}
  D.~St\"ockinger and P.~Varso,
  %``FeynArts model file for MSSM transition counterterms from DREG to DRED,''
  Comput.\ Phys.\ Commun.\  {\bf 183} (2012) 422
%  doi:10.1016/j.cpc.2011.10.010
  [arXiv:1109.6484 [hep-ph]].
  %%CITATION = doi:10.1016/j.cpc.2011.10.010;%%
  %6 citations counted in INSPIRE as of 28 Mar 2018


\bibitem{JJ}
I.~Jack and D.~R.~T.~Jones,
%``Regularisation of supersymmetric theories'', 
%in {\em Kane, G.L. (ed.): Perspectives on supersymmetry} 149-167;
[arXiv:hep-ph/9707278].
%%CITATION = HEP-PH 9707278;%%


\bibitem{DS05}
D.~St\"ockinger,
%``Regularization by dimensional reduction: Consistency, quantum action
%principle, and supersymmetry,''
{\em JHEP} {\bf 0503} (2005) 076,
hep-ph/0503129.
%%CITATION = HEP-PH 0503129;%%

  
%\cite{Harlander:2006xq}
\bibitem{Harlander:2006xq}
  R.~V.~Harlander, D.~R.~T.~Jones, P.~Kant, L.~Mihaila and M.~Steinhauser,
  %``Four-loop beta function and mass anomalous dimension in dimensional reduction,''
  JHEP {\bf 0612} (2006) 024
%  doi:10.1088/1126-6708/2006/12/024
  [hep-ph/0610206].
  %%CITATION = doi:10.1088/1126-6708/2006/12/024;%%
  %64 citations counted in INSPIRE as of 28 Mar 2018

  %\cite{Harlander:2009mn}
\bibitem{Harlander:2009mn}
  R.~V.~Harlander, L.~Mihaila and M.~Steinhauser,
  %``The SUSY-QCD beta function to three loops,''
  Eur.\ Phys.\ J.\ C {\bf 63} (2009) 383
%  doi:10.1140/epjc/s10052-009-1109-9
  [arXiv:0905.4807 [hep-ph]].
  %%CITATION = doi:10.1140/epjc/s10052-009-1109-9;%%
  %32 citations counted in INSPIRE as of 28 Mar 2018

  

  
  
\bibitem{BM}
               P.~Breitenlohner and D.~Maison,
               {\em Commun. Math. Phys.} {\bf 52} (1977) 11.
               %%CITATION = CMPHA,52,11;%%

              

%\cite{Jack:1994bn}
\bibitem{JJR}
  I.~Jack, D.~R.~T.~Jones and K.~L.~Roberts,
  %``Equivalence Of Dimensional Reduction And Dimensional Regularization,''
  Z.\ Phys.\  C {\bf 63}, 151 (1994),
%  [arXiv:hep-ph/9401349].
  %%CITATION = ZEPYA,C63,151;%%
%\cite{Jack:1993ws}
%\bibitem{Jack:1993ws}
%  I.~Jack, D.~R.~T.~Jones and K.~L.~Roberts,
  %``Dimensional Reduction In Nonsupersymmetric Theories,''
  Z.\ Phys.\  C {\bf 62}, 161 (1994).
%  [arXiv:hep-ph/9310301].
  %%CITATION = ZEPYA,C62,161;%%



\bibitem{IRstructure}
%\bibitem{ASDS05}
A.~Signer and D.~St\"ockinger, 
{\em Phys.\ Lett.}\ {\bf B 626} (2005) 127;
%hep-ph/0508203.
%%CITATION = HEP-PH 0508203;%%
%
%\cite{Signer:2008va}
%\bibitem{ASDS08}
%  A.~Signer and D.~St\"ockinger,
  %``Using Dimensional Reduction for Hadronic Collisions,''
  Nucl.\ Phys.\ B {\bf 808} (2009) 88
%  doi:10.1016/j.nuclphysb.2008.09.016
  [arXiv:0807.4424 [hep-ph]].
  %%CITATION = doi:10.1016/j.nuclphysb.2008.09.016;%%
  %46 citations counted in INSPIRE as of 28 Mar 2018
%
%\cite{Gnendiger:2014nxa}
%\bibitem{Gnendiger:2014nxa}
  C.~Gnendiger, A.~Signer and D.~Stöckinger,
  %``The infrared structure of QCD amplitudes and $H \to gg$ in FDH and DRED,''
  Phys.\ Lett.\ B {\bf 733} (2014) 296
%  doi:10.1016/j.physletb.2014.05.003
  [arXiv:1404.2171 [hep-ph]].
  %%CITATION = doi:10.1016/j.physletb.2014.05.003;%%
  %11 citations counted in INSPIRE as of 28 Mar 2018
%
%
%
%\cite{Broggio:2015ata}
%\bibitem{Broggio:2015ata}
  A.~Broggio, C.~Gnendiger, A.~Signer, D.~Stöckinger and A.~Visconti,
  %``Computation of $H\rightarrow gg$ in $\scriptsize{DRED}$ and $\scriptsize{FDH}$: renormalization, operator mixing, and explicit two-loop results,''
  Eur.\ Phys.\ J.\ C {\bf 75} (2015) no.9,  418
%  doi:10.1140/epjc/s10052-015-3619-y
  [arXiv:1503.09103 [hep-ph]];
  %%CITATION = doi:10.1140/epjc/s10052-015-3619-y;%%
  %7 citations counted in INSPIRE as of 28 Mar 2018
%\cite{Broggio:2015dga}
%\bibitem{Broggio:2015dga}
%  A.~Broggio, C.~Gnendiger, A.~Signer, D.~Stöckinger and A.~Visconti,
  %``SCET approach to regularization-scheme dependence of QCD amplitudes,''
  JHEP {\bf 1601} (2016) 078
%  doi:10.1007/JHEP01(2016)078
  [arXiv:1506.05301 [hep-ph]].
  %%CITATION = doi:10.1007/JHEP01(2016)078;%%
  %8 citations counted in INSPIRE as of 28 Mar 2018
%
%\bibitem{Kilgore:2012tb}
  W.\ Kilgore, %The four dimensional helicity scheme beyond one loop
  Phys.~Rev.~{\bf D 86} (2012) 014019 [arXiv:1205.4015].



\bibitem{multiloopepsilonscalars}
%\cite{Harlander:2006rj}
%\bibitem{Harlander:2006rj}
  R.~Harlander, P.~Kant, L.~Mihaila and M.~Steinhauser,
  %``Dimensional Reduction applied to QCD at three loops,''
  JHEP {\bf 0609} (2006) 053
%  doi:10.1088/1126-6708/2006/09/053
  [hep-ph/0607240].
  %%CITATION = doi:10.1088/1126-6708/2006/09/053;%%
  %50 citations counted in INSPIRE as of 28 Mar 2018
%
%\bibitem{Kilgore}
  W.~B.~Kilgore,
  %``Regularization Schemes and Higher Order Corrections,''
  Phys.\ Rev.\  {\bf D83 } (2011)  114005.
%  [arXiv:1102.5353 [hep-ph]].


\bibitem{HKRRSS}
W.~Hollik, E.~Kraus, M.~Roth, C.~Rupp, K.~Sibold and D.~St\"ockinger,
%``Renormalization of the minimal supersymmetric standard model,''
{\em Nucl.\ Phys.}{\bf\ B} {\bf 639} (2002) 3,
hep-ph/0204350.
%%CITATION = HEP-PH 0204350;%%

\bibitem{STIChecks2}
W.~Hollik and D.~St\"ockinger,
%``Regularization and supersymmetry-restoring counterterms in  supersymmetric
%QCD,''
{\em Eur.\ Phys.\ J.}{\bf\ C} {\bf 20} (2001) 105,
hep-ph/0103009.
%%CITATION = HEP-PH 0103009;%%
\bibitem{STIChecks3}
I.~Fischer, W.~Hollik, M.~Roth and D.~St\"ockinger,
%``Restoration of supersymmetric Slavnov-Taylor and Ward identities in  presence
%of soft and spontaneous symmetry breaking,''
{\em Phys.\ Rev.}{\bf\ D} {\bf 69} (2004) 015004,
hep-ph/0310191.
%%CITATION = HEP-PH 0310191;%%

%% \bibitem{betachecks}
%% I.~Jack, D.~R.~T.~Jones and C.~G.~North,
%% %``$N=1$ supersymmetry and the three loop anomalous dimension for the chiral
%% %superfield,''
%% {\em Nucl.\ Phys.}{\bf\ B} {\bf 473} (1996) 308,
%% hep-ph/9603386;
%% %%CITATION = HEP-PH 9603386;%%
%% %``N = 1 supersymmetry and the three loop gauge beta function,''
%% {\em Phys.\ Lett.}{\bf\ B} {\bf 386} (1996) 138,
%% hep-ph/9606323;
%% %%CITATION = HEP-PH 9606323;%%
%% %``Scheme dependence and the NSVZ beta-function,''
%% {\em Nucl.\ Phys.}{\bf\ B} {\bf 486} (1997) 479,
%% hep-ph/9609325.
%% %%CITATION = HEP-PH 9609325;%%



%% \bibitem{SSTI1}
%% P.~L.~White,
%% %``An Analysis of the cohomology structure of superYang-Mills coupled to
%% %matter,''
%% {\em Class.\ Quant.\ Grav.}\  {\bf 9} (1992) 1663.
%% %%CITATION = CQGRD,9,1663;%%

%% \bibitem{SSTI2}
%% N.~Maggiore, O.~Piguet and S.~Wolf,
%% %``Algebraic renormalization of N=1 supersymmetric gauge theories,''
%% {\em Nucl.\ Phys.}{\bf\ B }{\bf 458} (1996) 403
%% [Erratum-ibid.\ B {\bf 469} (1996) 513],
%% hep-th/9507045;
%% %%CITATION = HEP-TH 9507045;%%
%% %``Algebraic Renormalization of $N=1$ Supersymmetric Gauge Theories with
%% %Supersymmetry Breaking Masses,''
%% {\em Nucl.\ Phys.}{\bf\ B }{\bf 476} (1996) 329,
%% hep-th/9604002.
%% %%CITATION = HEP-TH 9604002;%%

%% \bibitem{SSTIus1}
%% W.~Hollik, E.~Kraus and D.~St\"ockinger,
%% %``Renormalization of supersymmetric Yang-Mills theories with soft
%% %supersymmetry breaking,''
%% {\em Eur.\ Phys.\ J.}{\bf\ C }{\bf 23} (2002) 735,
%% hep-ph/0007134.
%% %%CITATION = HEP-PH 0007134;%%

%% \bibitem{Siegel80}
%% W.~Siegel,
%% %``Inconsistency Of Supersymmetric Dimensional Regularization,''
%% {\em Phys.\ Lett.}{\bf\ B }{\bf 94} (1980) 37.
%% %%CITATION = PHLTA,B94,37;%%




%% %%%%%%%%%%%%%%%%%%%%%%%%%%%%%%%%%%%%% not required?

%% \bibitem{LEPBound}
%% G.~Abbiendi  et al.  [ALEPH, DELPHI, L3, OPAL Collaborations and LEP
%%   Working Group for Higgs boson searches],
%% %``Search for the standard model Higgs boson at LEP,''
%% {\em Phys.\ Lett.}{\bf\ B} {\bf 565}, 61 (2003),
%% hep-ex/0306033.
%% %%CITATION = HEP-EX 0306033;%%
%% \bibitem{LHCUncertainty}
%% J.\ Branson et al.~[CMS Collaboration], hep-ph/0110021.

%% \bibitem{ILCUncertainty}
%%  J.~Aguilar-Saavedra et al.,
%%                 TESLA TDR Part~3: 
%%                 ``Physics at an $e^+e^-$ Linear Collider'', 
%%                 hep-ph/0106315,
%%                 %%CITATION = HEP-PH 0106315;%%
%%                 see: {\tt tesla.desy.de/tdr/} ;\\
%%               T.~Abe et al.
%%                      [American Linear Collider Working Group Collaboration],
%%                      {\it Resource book for Snowmass 2001}, 
%%                      hep-ex/0106056;\\
%%                      %%CITATION = HEP-EX 0106056;%%
%%               K.~Abe et al. 
%%                   [ACFA Linear Collider Working Group Collaboration],
%%                   hep-ph/0109166.
%%                   %%CITATION = HEP-PH 0109166;%%

%% \bibitem{Review}
%% S.~Heinemeyer, W.~Hollik and G.~Weiglein,
%% %``Electroweak precision observables in the minimal supersymmetric  standard
%% %model,''
%% hep-ph/0412214.
%% %%CITATION = HEP-PH 0412214;%%


%% \bibitem{DREDProc}
%% D.~St\"ockinger, {\em Proceedings of the 2005 International Linear
%%   Collider Workshop (LCWS 2005), Stanford, California},
%% %``Regularization of supersymmetric theories: Recent improvements,''
%% hep-ph/0506258.
%% %%CITATION = HEP-PH 0506258;%%


%% \bibitem{Harlander05}
%% R.~V.~Harlander and F.~Hofmann,
%% %``Pseudo-scalar Higgs production at next-to-leading order SUSY-QCD,''
%% hep-ph/0507041.
%% %%CITATION = HEP-PH 0507041;%%

\end{thebibliography}
\end{document}